\documentclass[useAMS,usenatbib]{mn2e}
\usepackage{graphicx,subfigure,ram}

\begin{document}

\title[H{\sc i} Study of LSB galaxies]{GMRT H{\sc i}  study of
giant low surface brightness galaxies}
\author[Mishra et. al.]{Alka Mishra$^{1,2}$\thanks{E-mail: alkam7@gmail.com (AM), 
      nkprasadnetra@gmail.com (NGK)},
             N.~G.~Kantharia,$^{3\star}$, 
             M.~Das$^{4}$,   
             A.~Omar$^{1}$ and
             D.~C.~Srivastava $^{2}$ \\
$^{1}$Aryabhatta Research Institute of Observational Sciences (ARIES), Manora Peak, Nainital 263 002,India\\
$^{2}$Department of Physics, D.D.U. Gorakhpur University,Gorakhpur 273 009,India\\
$^{3}$National Centre for Radio Astrophysics,TIFR ,Pune 411 007, India\\
$^{4}$Indian Institute of Astrophysics, Koramangala,Bangalore 560 034,India\\}

\date{}

\maketitle

\begin{abstract}

We present H{\sc i} observations of four giant low surface brightness 
(GLSB) galaxies  UGC~1378, UGC~1922, UGC~4422 and UM~163 using the 
Giant Meterwave Radio Telescope (GMRT).
We include H{\sc i} results on UGC~2936, UGC~6614 and Malin~2 from 
literature.  H{\sc i} is detected from all the galaxies and the extent 
is roughly twice the optical size; in UM 163, H{\sc i} is detected along 
a broken disk encircling the optical galaxy. We combine our results with 
those in literature to further understand these systems.   
The main results are the following: (1) The peak H{\sc i} surface 
densities in GLSB galaxies are 
several times $10^{21}$ cm$^{-2}$. The H{\sc i} mass is between 
$0.3-4 \times 10^{10}$ M$_\odot$, dynamical mass ranges 
from a few times $10^{11}$ M$_\odot$ to a few times $10^{12}$ M$_\odot$.  
(2) The rotation curves of GLSB galaxies are flat to the outermost 
measured point with rotation velocities of the seven GLSB galaxies being
between  225 and 432 km s$^{-1}$. 
(3) Recent star formation traced by near-ultraviolet emission in 
five GLSB galaxies in our sample appears to be located in rings around the galaxy 
centre.  We suggest that this could be due to a stochastic burst of star formation at 
one location in the galaxy being propagated along a ring over a rotation period. 
(4) The H{\sc i} is correlated with recent star formation in five 
of the seven GLSB galaxies.  
\end{abstract}


\section{Introduction}

In the last few decades low surface brightness (LSB) galaxies 
have received a lot of attention.  LSB galaxies are
defined as disk galaxies with a central surface brightness $\mu_{B}$ $<$ 23 
mag arcsec$^{-2}$ \citep{impey97} 
which distinguishes them from the high surface brightness (HSB) galaxies. 
The low surface brightness galaxies are missing in earlier optical catalogues
due to sky brightness causing a strong bias against detection of objects 
with central surface brightness fainter than the Freeman 
value ($\mu_{B}$ $=$ 21.65 $\pm$ 0.35 mag arcsec$^{-2}$). 
The accidental discovery of the first giant LSB (GLSB) galaxy Malin-1 by \cite{bothun87}  
demonstrates that there might be a large number of galaxies with similar 
properties  (\citealt{bothun90}; 
\citealt{sprayberry93}; \citealt{sprayberry95}; \citealt{neil2000}).
Many LSB galaxies have since been identified 
in deep and large field survey (\citealt{schombert88};
\citealt{schombert92}; \citealt{davies94}; 
\citealt{schwartzenberg95}; \citealt{sprayberry95}). 
Surveys have revealed some LSB galaxies  having  
disk scale lengths comparable to the Milky Way. 
LSB galaxies are host to a range of stellar populations,  
covering the entire range of H-R diagram (\citealt{zackrisson05}; 
\citealt{zhong08}).  LSB galaxies show deficit in molecular
gas \citep{schombert90}  which has been detected in only a few galaxies 
(\citealt{das06}; \citealt{oneil00}; \citealt{das10}). LSB galaxies are 
very rich in H{\sc i} but have low gas surface densities and slowly 
rising H{\sc i} rotation curves. They have higher M$_{H{\sc I}}$/L$_{B}$ 
ratios as compared to HSB galaxies \citep{deblok96a}. 
The H{\sc i} masses are close to  10$^{9}$ M$_{\odot}$ \citep{deblok96b} and 
the H{\sc i} surface density are usually close to the critical density for 
star formation (\citealt{kennicutt89}; \citealt{vanderhulst93}; 
\citealt{deblok96a}). LSB galaxies comprise a significant fraction of the spiral galaxy 
population \citep{mcgaugh95} and hence form an important class of spiral galaxies. 
Some LSB spirals have  been found to be larger than the  normal galaxy - 
these have been referred to as Giant LSB galaxies.  
In a H{\sc i} survey of 116 LSB galaxies, selected from the \cite{bothun85} 
catalog, carried out using the Arecibo and Nancay telescopes, 81 LSB galaxies were 
detected out of which 38 have H{\sc i} mass $>10^{10}$ M$_\odot$ \citep{oneil04}.  
These 38 galaxies might be GLSB galaxies but 
this survey was done with a large telescope beam; interferometric observations
are required to  distinguish GLSB galaxies from HSB galaxies in the field of view.

On the basis of the study to date, it is clear that LSB galaxies have some
different properties compared  to the HSB  galaxies  possibly due to different 
evolutionary sequence. The galaxy 
evolution and its star formation history are strongly correlated with 
the environment in which the galaxy is evolving. The effect of the 
environment upon bright galaxies are well studied and there are noticeable 
variations in galaxy properties like morphology, color and  magnitude.

In the present paper, we take up H{\sc i} 21cm-line study of the sample 
of seven GLSB galaxies, UGC~1378, UGC~1922, UGC~4422, UM~163, UGC~2936, UGC~6614 
and Malin~2. This sample  has been studied in the radio continuum by us 
in an earlier paper, \cite{mishra15}. Here, we present H{\sc  i} 21cm-line 
observations of the first four galaxies i. e. UGC~1378, UGC~1922, UGC~4422, UM~163 
and employ the results on UGC~6614, Malin~2 from \cite{pickering97} and for 
UGC~2936 by  \cite{pickering99}. We refer to the 
basic properties and the parameters of the galaxies of the entire sample 
as given in Table 1 of  \cite{mishra15}.

\begin{figure*}
\vspace*{50pt}
\subfigure[]{\includegraphics[height=7.0cm,angle=-90]{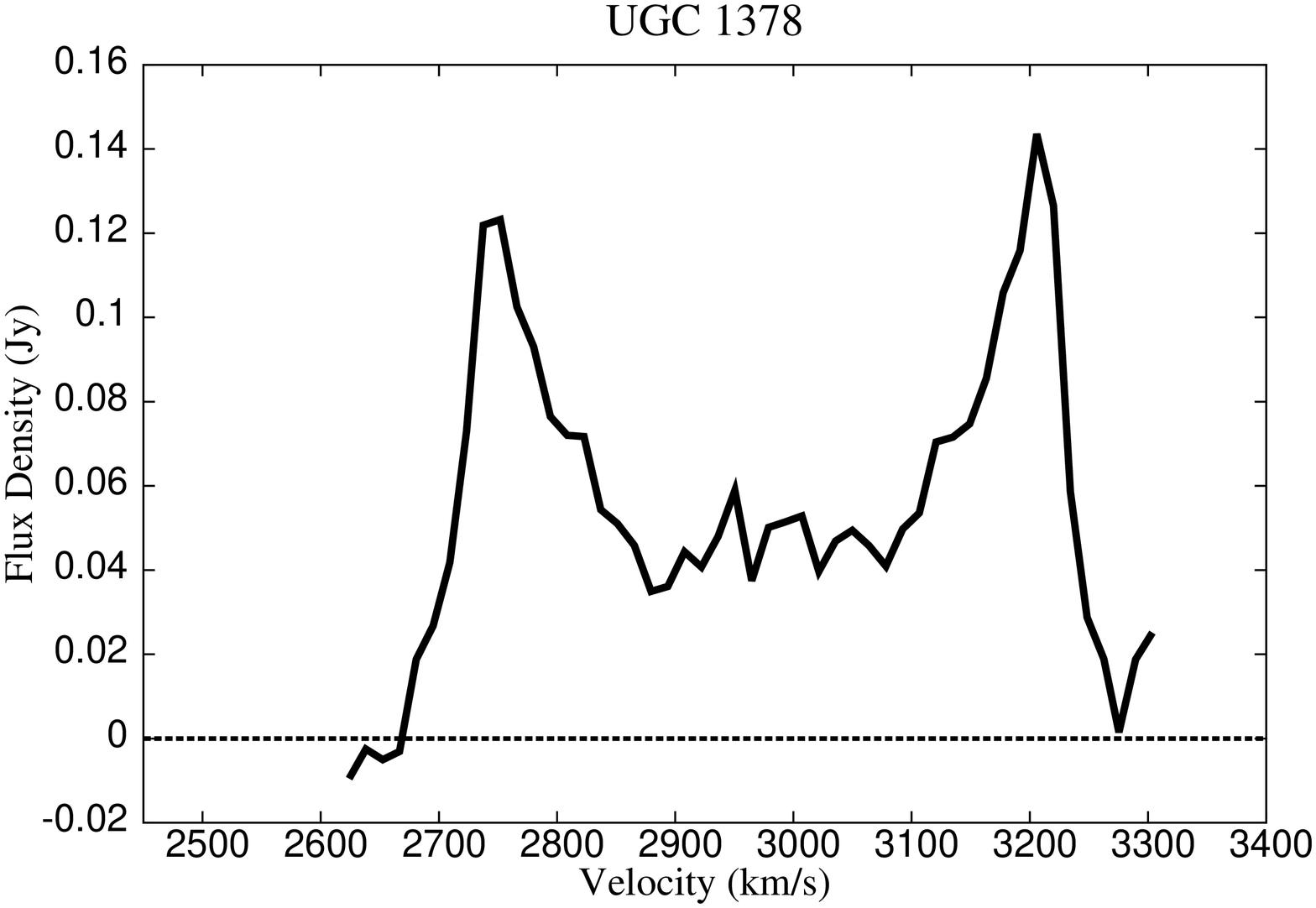}}
\subfigure[]{\includegraphics[height=7.0cm,angle=-90]{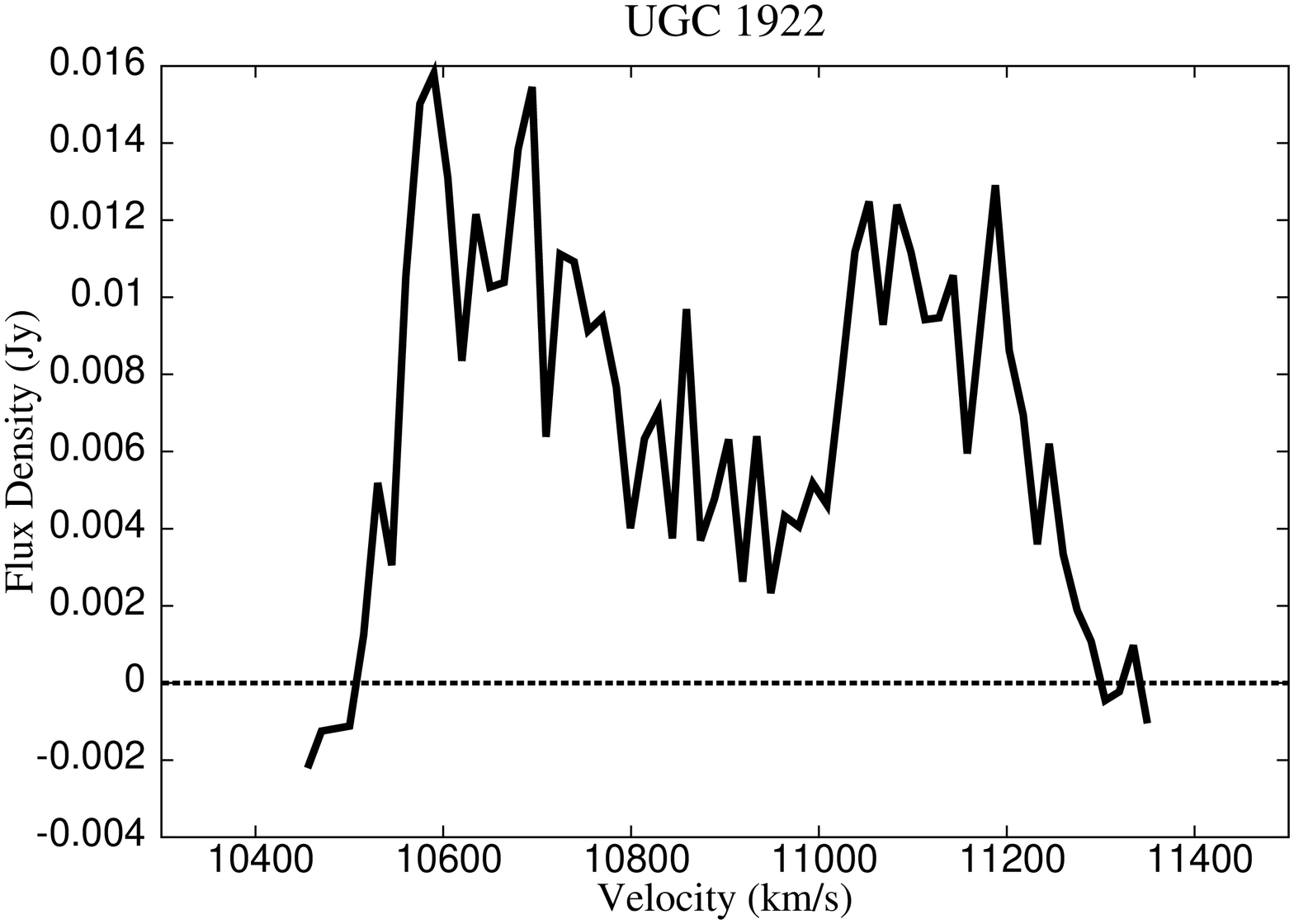}}
\subfigure[]{\includegraphics[height=7.0cm,angle=-90]{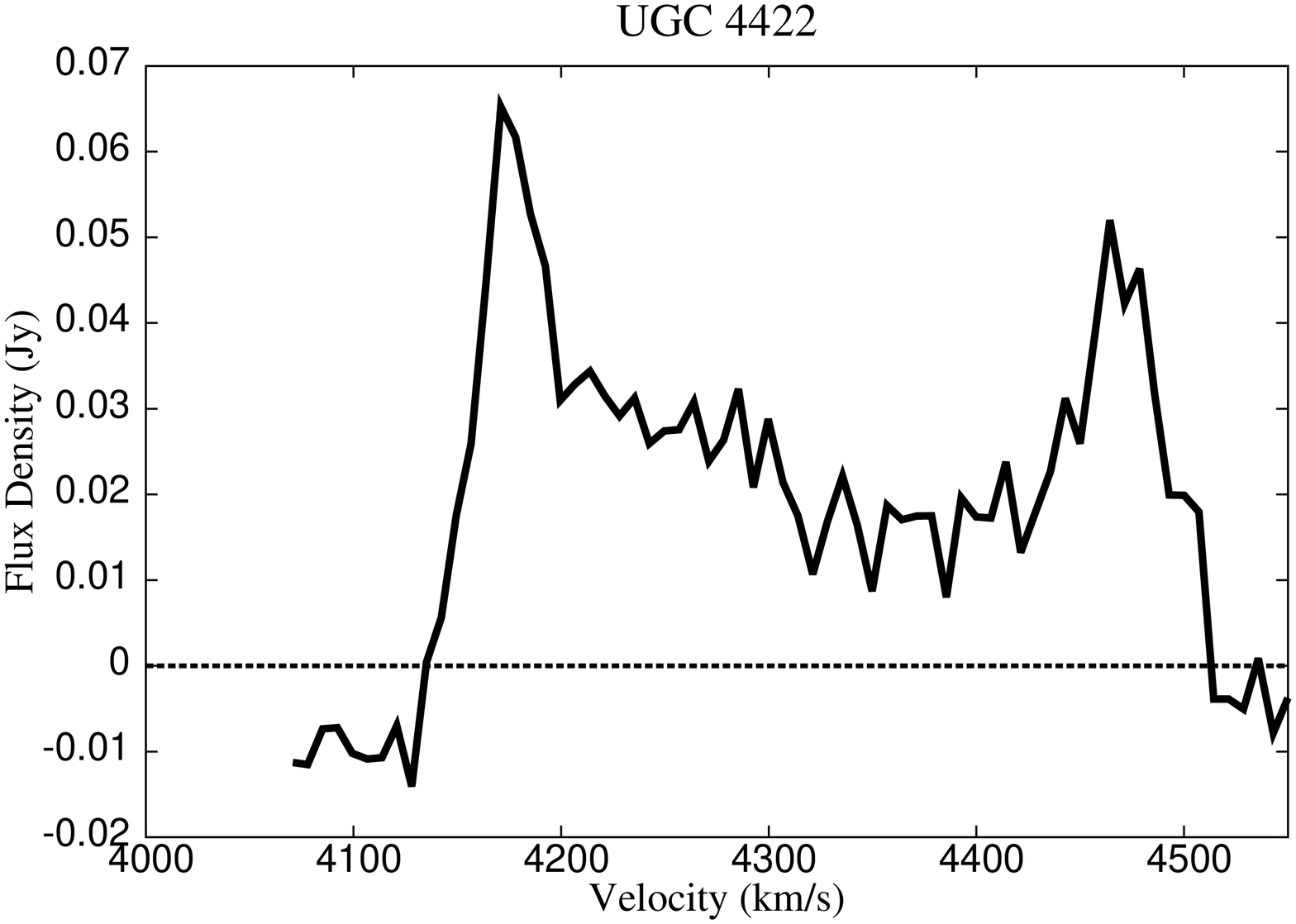}}
\subfigure[]{\includegraphics[height=7.0cm,angle=-90]{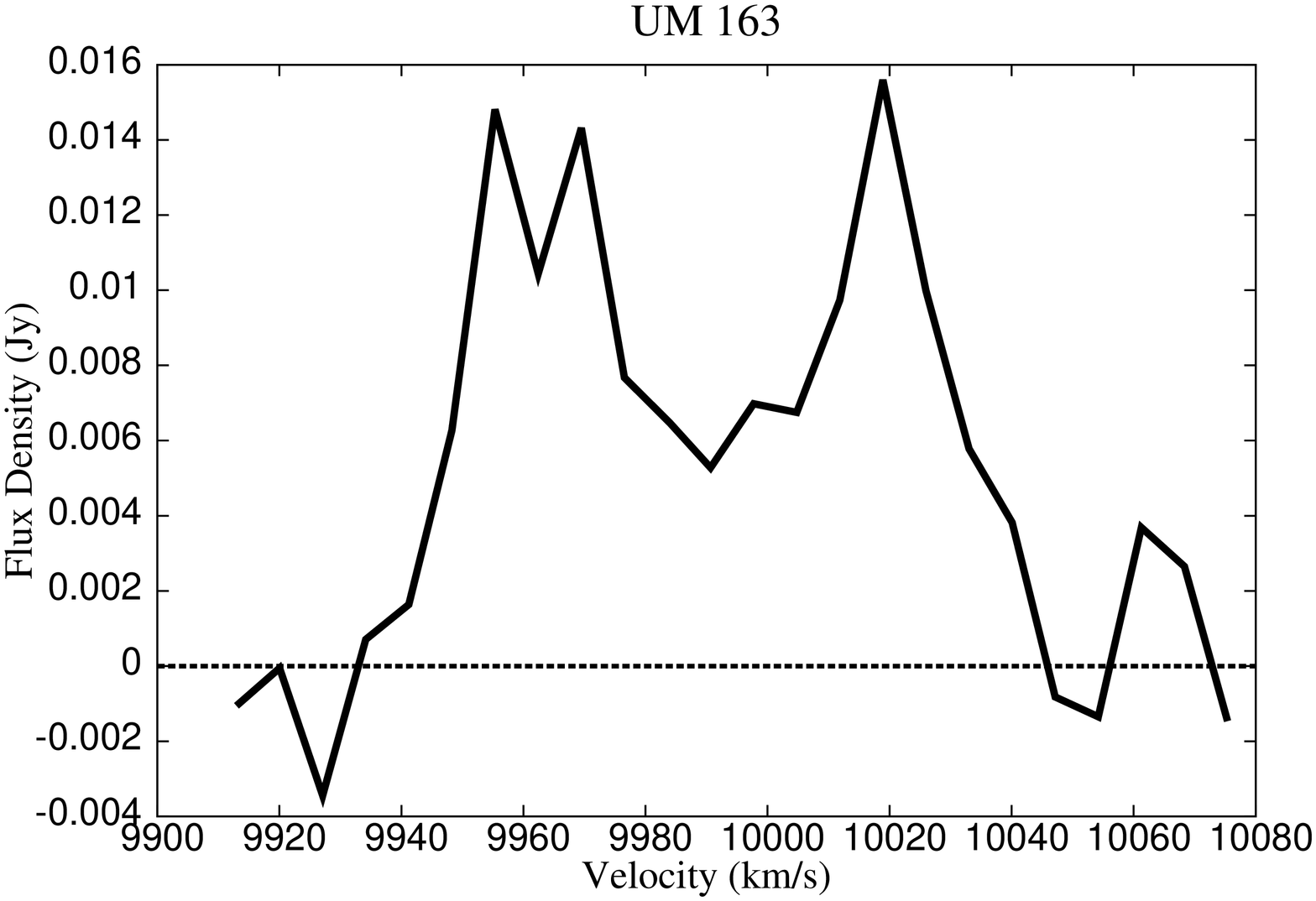}}
\caption{Global H{\sc i} emission profiles made from the low resolution maps:  
(a) {\bf UGC 1378}, (b) {\bf UGC 1922}, (c) {\bf UGC 4422} and  
(d) {\bf UM 163} showing the double-humped profile, characteristic of rotating disks.}
\label{fig:figure.1}
\end{figure*}

\begin{table*}
 \centering
\begin{minipage}{100mm}
  \caption{GMRT Observations}
    \begin{tabular}{@{}lllll@{}}\hline \hline

\footnotetext[1] {Flux densities from SETJY}
\footnotetext[2] {Flux densities and error from GETJY} 
\footnotetext[3] {These data were taken with the old hardware backend at GMRT}

                   & UGC~1378 &  UGC~1922 & UGC~4422 & UM~163   \\ \hline \hline
Observing dates        & 30-12-11 &  31-12-11 & 29-12-11 & 28-07-06  \\
Flux calibrator(s)  & 3C~48     &  3C~48   &  3C~147, 3C~286  &  3C~48, 3C~286  \\
~~Flux density(Jy)$^{a}$    &  16.08    & 16.41  & 22.27, 14.86   &  16.31, 14.96 \\
Phase calibrator    & 0217+738 &  3C 48   & 0842+185       &  2225-049   \\
~~Flux density(Jy)$^{b}$& 2.60$\pm$0.03&16.41$\pm$0.02 &1.16$\pm$0.01 & 7.53$\pm$0.15 \\
Central frequency (MHz) &  1397     &  1362  &    1392        &     1372       \\
Central velocity (kms$^{-1}$) &  2935     &  10894  &    4330      &     10022       \\
Bandwidth  (MHz)  &   16       &   16     &   16         &  04   \\
t$_{source}$ (hr) &     5.5    &  6.0        &  5.5         &    5.5   \\
No. of channels   &      256    &    256      &  512        &  128     \\
Channel resolution (km s$^{-1}$) &  13.8   &   14.2      &  6.9        &  6.6      \\
Synthesized beam/   & $50^{\prime\prime}\times 50^{\prime\prime}$ & $45^{\prime\prime}\times 45^{\prime\prime}$ & 
$45^{\prime\prime}\times 45^{\prime\prime}$  & $45^{\prime\prime}\times 45^{\prime\prime}$\\
Ang resolution&$25^{\prime\prime}\times 25^{\prime\prime}$ & $30^{\prime\prime}\times 30^{\prime\prime}$ & 
$30^{\prime\prime}\times 30^{\prime\prime}$  & $32^{\prime\prime}\times 32^{\prime\prime}$ \\
&$15^{\prime\prime}\times 15^{\prime\prime}$ & $15^{\prime\prime}\times 15^{\prime\prime}$& 
$15^{\prime\prime}\times 15^{\prime\prime}$ &  $20^{\prime\prime}\times 20^{\prime\prime}$ \\
RMS noise per channel  & 1.5, 1.3, 1.0 &  1.7, 1.5, 1.1 & 2.4, 2.0, 1.5 &   1.1, 1.0, 0.8$^{c}$\\
at diff. res.(mJy/beam) &              &          &          &           \\ 
\hline \hline
\end{tabular}
\end{minipage}
\label{tab:lsb_obser.hi}
\end{table*}

\section{Observations And Data Analysis} 
\footnotetext[1]{AIPS is distributed by the National Radio 
Astronomy Observatory, which is a facility of the National 
Science Foundation operated under cooperative
agreement by Associated Universities, Inc.}

The observations were made with the Giant Metrewave 
Radio Telescope (GMRT;  \citealt{swarup91}). 
The GMRT is an array of 30 antennas, each of 45 m in diameter spread over a 
maximum baseline length of 25 km.  At a frequency of $\sim$ 1420 MHz, 
the system temperature and gain of the 
instrument are listed as 76 K and 0.22 KJy$^{-1}$ respectively. 
The data on UM~163 were taken with the old hardware backend at 
GMRT and we had some flux calibration issues.  The maps presented here have
been made with the best possible solutions but we would like to caution the 
reader that these maps and estimated column densities need to be 
confirmed by future observations. Flux calibration was carried out
using scans on any of the available standard calibrators (3C~48/ 3C~147/ 3C~286) 
which were observed at the start, end or middle of the observing run. 
Phase calibration was carried out using a nearby  calibrator source. 
Phase calibrator was observed once in every 30 minutes. Bandpass calibration was
carried out using the flux calibrators. The data were reduced
using standard tasks in NRAO AIPS$^{1}$.  Bad data due to dead
antennas and those with significantly lower gain
than others, and radio frequency interference (RFI)
were flagged. After removing bad data, the final gain
tables were generated and the target source data
were calibrated. The data were then imaged by using
a cell size of $1^{\prime\prime} \times 1^{\prime\prime}$.
The spectral  line cube was made after  removing the continuum
emission  using the AIPS tasks UVSUB and IMLIN. The task IMAGR 
was  used to get the final three-dimensional deconvolved H{\sc i} 
data cubes. We applied  primary beam correction to all images by using  
PBCOR. DOPSET was used to obtain the frequency offset for the 
heliocentric velocity of the galaxy.
From these cubes, the  H{\sc i} emission  and
the H{\sc i} velocity field were extracted using the AIPS
task MOMNT. Data cubes were  made at various cutoffs in the uv plane, 
including 0-20 k$\lambda$, 0-10 k$\lambda$ and 0-05 k$\lambda$ and these
were then convolved to a circular beam.  These resulted in respective angular
resolutions  of about $15^{\prime\prime}\times 15^{\prime\prime}$,
$25^{\prime\prime}\times 25^{\prime\prime}$ and
$50^{\prime\prime}\times 50^{\prime\prime}$.  Details of
the observations  are summarized in Table \ref{tab:lsb_obser.hi}.

\begin{table*}
\centering
\begin{minipage}{200mm}
\caption{Properties of Observed H{\sc i}}
\begin{tabular}{@{}llllllll|lll@{}}\hline
\footnotetext[1] {Observed H{\sc i} linewidth}  
\footnotetext[2] {Average value from rotation curve fitting using tilted ring model. In case of UGC~1922 we 
derived the inclination from the observed H{\sc i} ellipticity}
\footnotetext[3]{H{\sc i} mass: M$_{HI}$= 2.36 $\times$ 10$^{5}$ D$^{2}$
                 $\int$ H{\sc i} flux, where the luminosity distance D is given in Mpc}
\footnotetext[4]{Dynamical mass M$_{dyn}$ = (V$_{rot}$)$^{2}$ R/G}
\footnotetext[5]{Single dish H{\sc i} flux from the GBT (beam $\sim$ 9$^{\prime}$)  \citep{hogg07}} 
\footnotetext[6]{Single dish H{\sc i} flux from the Arecibo (beam $\sim$ 3$^{\prime}$.3)  \citep{springob05}}
\footnotetext[7]{Single dish H{\sc i} flux from the Parkes (beam $\sim$ 14$^{\prime}$.3) \citep{doyle05}}
\footnotetext {* from Pickering et al. (1997, 1999)}

                                 &UGC~1378 & UGC~1922 & UGC~4422 & UM~163 & UGC~2936* & UGC~6614* & Malin~2*\\  \hline
H{\sc i} flux (Jy km s$^{-1}$)  & 34.5$\pm$2.2& 6.0$\pm$1.1 & 16.4$\pm$1.4 & 0.8$\pm$0.1 
& 13.6$\pm$1.7* & 15.0$\pm$0.8* & 4.4$\pm$0.3*   \\
W$_{50,obs}$  (km s$^{-1})$$^{a}$ & 500$\pm$15 & 640$\pm$14 & 320$\pm$8 &100$\pm$12 &500$\pm$10* &287$\pm$10* & 392$\pm$15* \\
V$_{sys}$  (km s$^{-1})$$^{b}$   & 2938$\pm$10&10853$\pm$12& 4308$\pm$10  & ....  & 3813$\pm$3* & 6353$\pm$7* &13835$\pm$5* \\
Inclination($^{\circ}$) $^{b}$   &  59$\pm$5  &  51$\pm$2  &  40$\pm$4      & \ldots.   & \ldots. & \ldots.& \ldots.   \\
Position angle ($^{\circ}$) $^{b}$ & 181$\pm$6 &  128$\pm$3 & 33$\pm$2   & \ldots.   & \ldots. & \ldots.&\ldots.    \\
V$_{rot}$ (km s$^{-1})$$^{b}$    &  282$\pm$11 &  432$\pm$12  &  254$\pm$8  & \ldots.   &  256* & 227* & 320*  \\
H{\sc i}(10$^{10}$ M$_{\odot}$)$^{c}$  &  1.2$\pm$0.2  &   3.2$\pm$0.4 &1.5$\pm$0.5  & 0.34$\pm$0.04   &  0.8$\pm$1.0* 
& 2.5$\pm$0.2* &3.6$\pm$0.4*  \\
M$_{dyn}$ (10$^{11}$ M$_{\odot}$)$^{d}$ & 7.4$\pm$0.9 & 22.0$\pm$2.1 & 4.5$\pm$1.0 & \ldots.& 5.1$\pm$1.2 &6.7$\pm$1.4& 25.5$\pm$2.8 \\
M$_{HI}$/ M$_{dyn}$ ($\%$)              &   1.6  &  1.4    &   3.3   & .... & 1.6  & 3.7  & 1.4     \\
half H{\sc i} mass radius (kpc) & 18.8 &  26  &  15.5   & .... & .... & .... & ....   \\
half light radius (kpc)         &  9.5 & 22.7 &   7.5   & .... & .... & .... & ....   \\
Single dish H{\sc i} flux (Jy km s$^{-1}$) & 35.8$^{e}$   & 9.2$^{f}$&  20.5$^{f}$  & 4.4 $^{g}$ &....  &....  &.... \\
Absolute B magnitude & -21.5 & -21.26 & -21.33 & -21.8  & -21.76 & -20.71 &  -21.58 \\
(from Hyperleda) & & & & & & & \\ \hline
\end{tabular}
\end{minipage}
\label{tab:lsb_hiresult}
\end{table*}

\section{Results}

\footnotetext[2]{http://skyview.gsfc.nasa.gov/current/cgi/query.pl}

The global H{\sc i} emission profiles of the sample galaxies  made 
from low resolution maps are displayed in Figures \ref {fig:figure.1}(a)-(d) 
and various parameters derived from the H{\sc i} data are listed in 
Table \ref{tab:lsb_hiresult}. The half light radius of the galaxies 
approximated by half of the semi-major axis as listed  by 
\cite{devaucouleurs91} and radius of the region enclosing half the  
H{\sc i} mass of the galaxy are listed in Table \ref{tab:lsb_hiresult}.  
While for UGC~1378 and UGC~4422 the half H{\sc i} mass 
radius is close to twice that of half the semi-major axis; 
in the case of UGC~1922, half H{\sc i} mass radius is comparable to 
half the semi-major axis. 
Double horn profiles, characteristic of a rotating disk are 
observed.  Total H{\sc i} intensity maps overlaid on near-infrared (NIR) 
from 2MASS$^{2}$,  B band emission from the Digitized Sky Survey (DSS)$^{2}$, 
near-ultraviolet (NUV) from GALEX$^{2}$, and H{\sc i} column density (moment 0) 
grey scale images are displayed in Figures \ref{fig:figure.2} - \ref{fig:figure.5}(a)-(d).  
H{\sc i} velocity (moment 1) and velocity dispersion (moment 2) maps of these galaxies are 
displayed  in Figures \ref{fig:figure.2}-\ref{fig:figure.5}(e)-(f).  
High dispersion regions correspond to the high H{\sc i} column density regions in the galaxies. 
The H{\sc i} disks of three galaxies; UGC~1378,  UGC~1922 and
UGC~4422 show  extended H{\sc i} disks of extent about twice the  
stellar disk whereas in UM~163, the H{\sc i} is detected in a broken disk 
around the optical galaxy. All the sample galaxies show mildly 
lopsided H{\sc i} distribution (e.g. PGC045080; \citealt{das07}) and a rotating disk.
All the galaxies also show H{\sc i} absorption in the centre due to the 
active nucleus in these galaxies.  
Figures \ref{fig:figure.6}(a)-(d) show zoomed-in H{\sc i} total
intensity  contours  overlaid on 610 MHz radio continuum  maps (grey scale) taken
from \cite{mishra15}.
Figures \ref{fig:figure.7}(a)-(c) show the rotation curves
estimated for UGC~1378, UGC~1922 and UGC~4422.  We could not estimate 
the rotation curve for UM~163 since the H{\sc i} is along a  broken disk. 
In Table \ref{tab:lsb_hiresult}, we also include the properties of the galaxies 
UGC~2936, UGC~6614 and Malin~2 taken from Pickering et al. (1997, 1999) 
to enable us to study all the seven galaxies whose radio continuum study 
was presented in \cite{mishra15}. In the following subsections we discuss 
the results on the individual galaxies.

\subsection{UGC~1378}
\begin{figure*}
\vspace*{50pt}
\subfigure[]{\includegraphics[height=8.0cm]{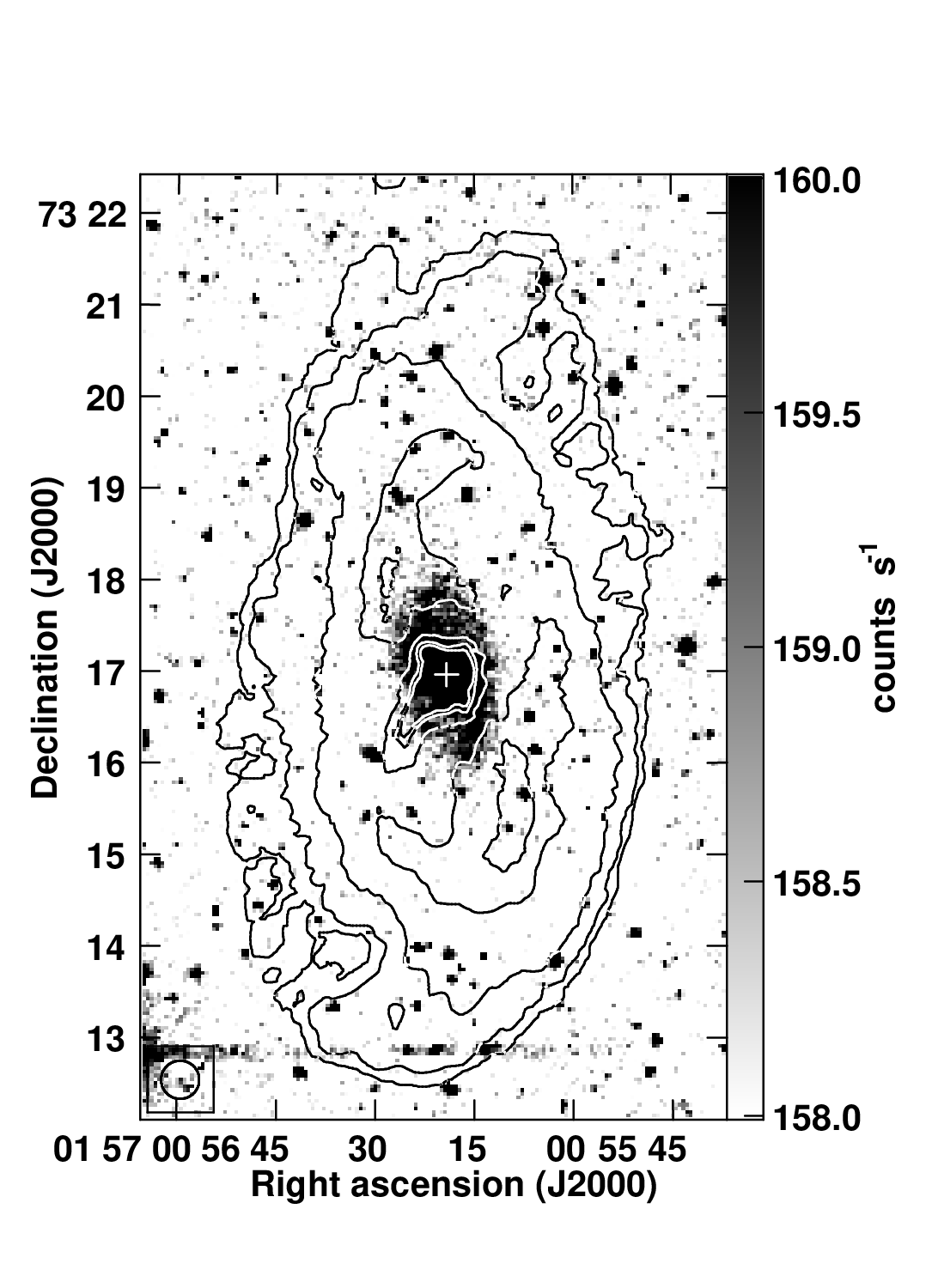}}
\subfigure[]{\includegraphics[height=8.0cm]{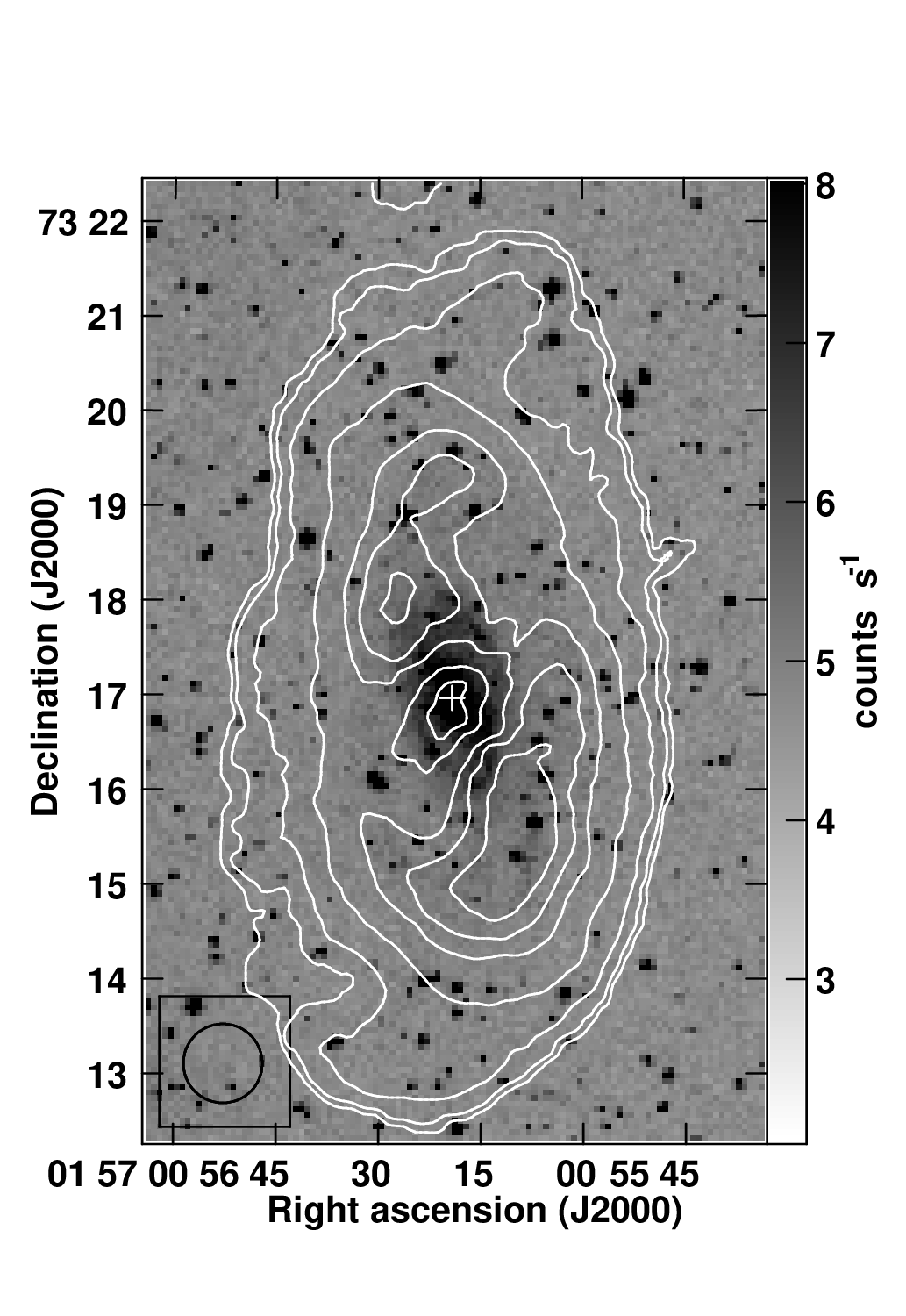}}
\subfigure[]{\includegraphics[height=8.0cm]{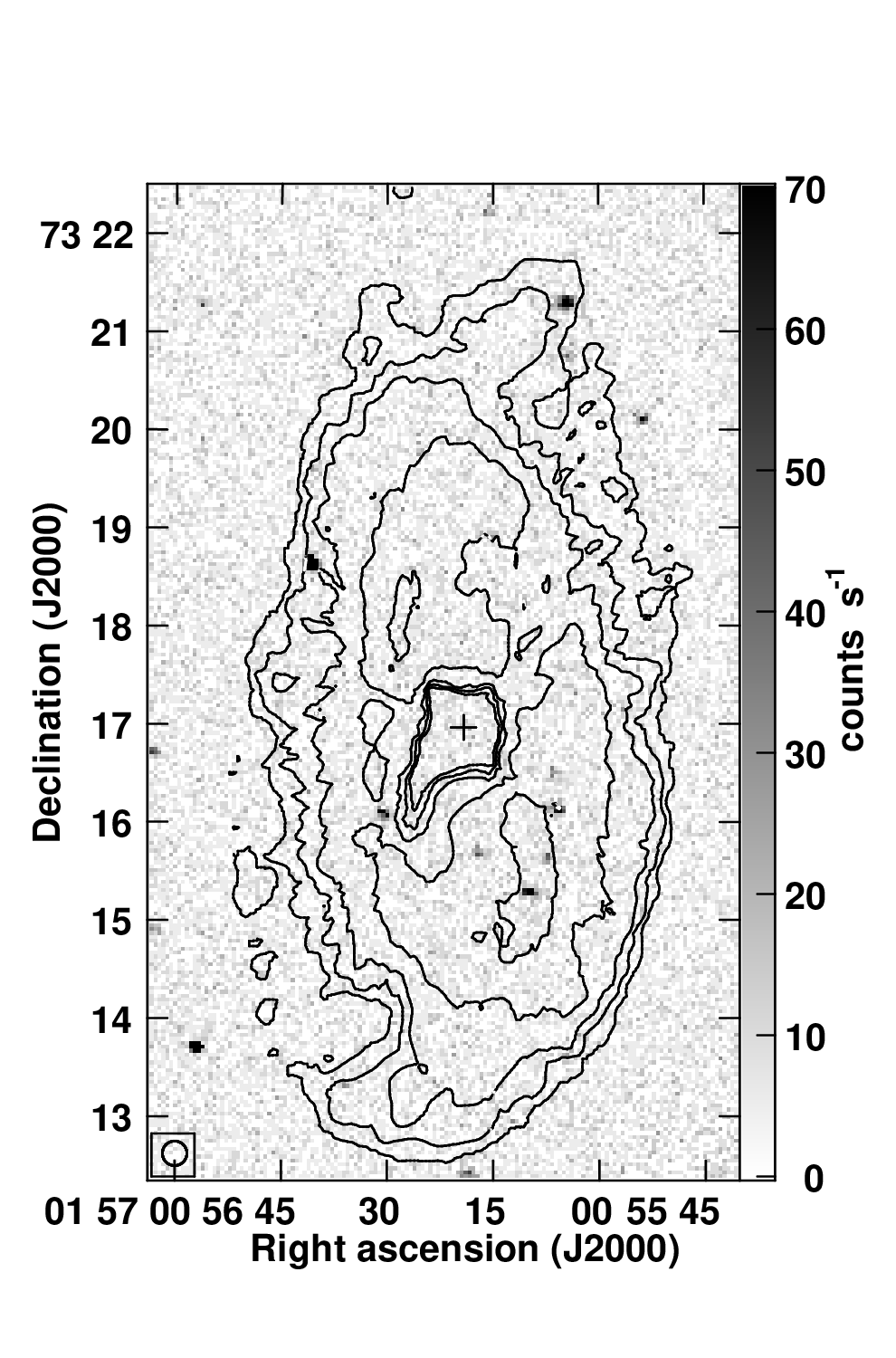}}
\subfigure[]{\includegraphics[height=8.0cm]{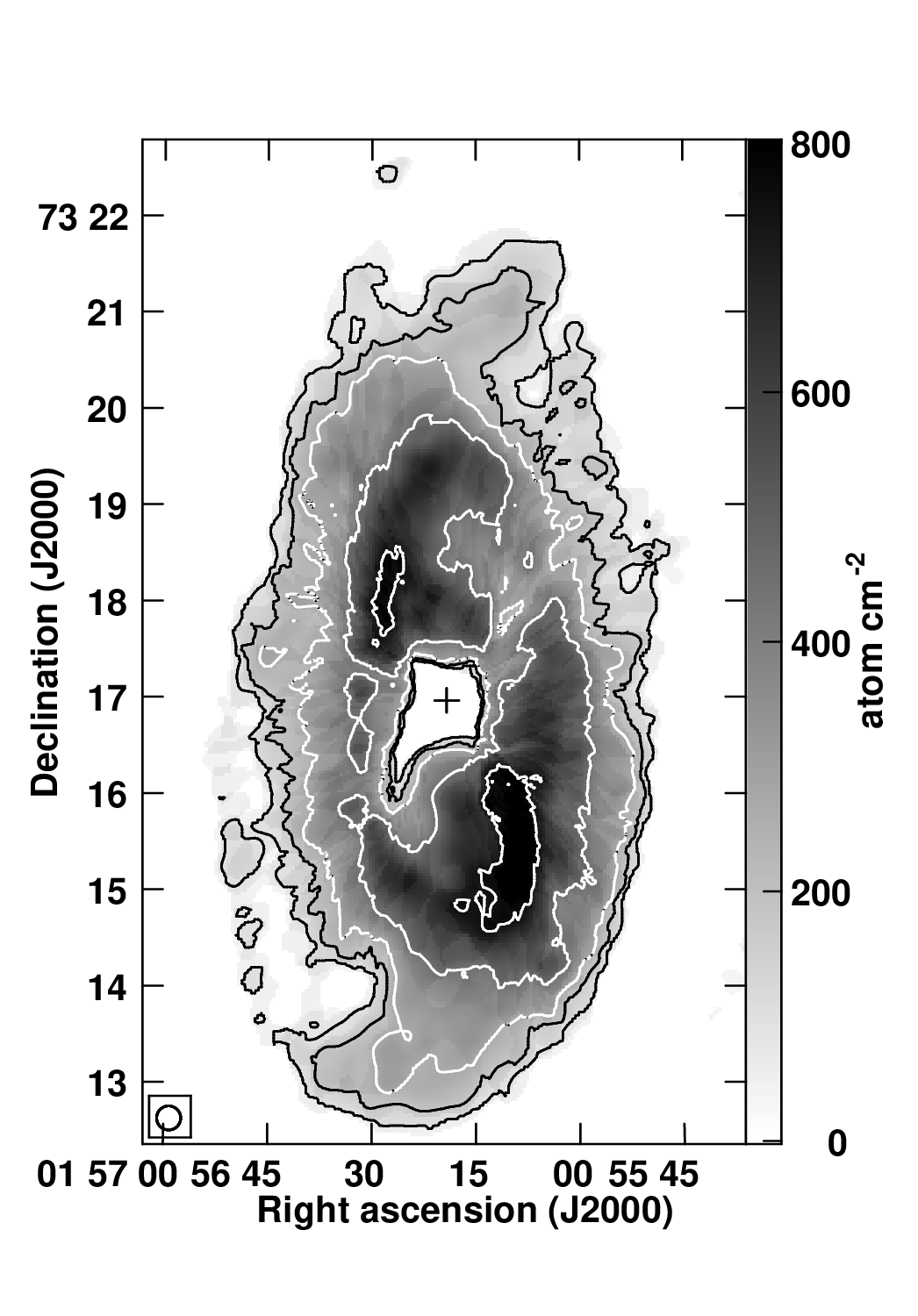}}
\subfigure[]{\includegraphics[height=8.0cm]{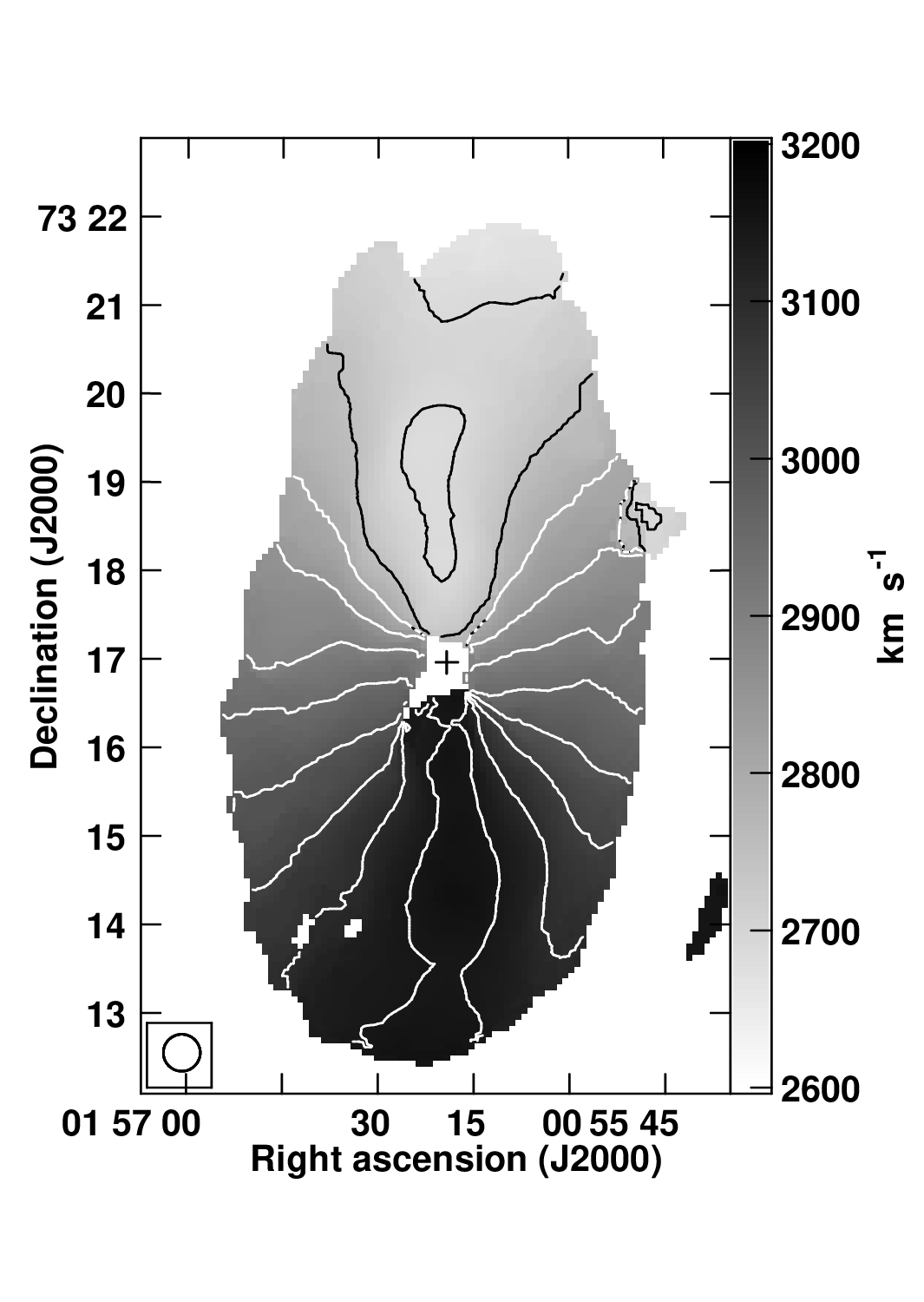}}
\subfigure[]{\includegraphics[height=8.0cm]{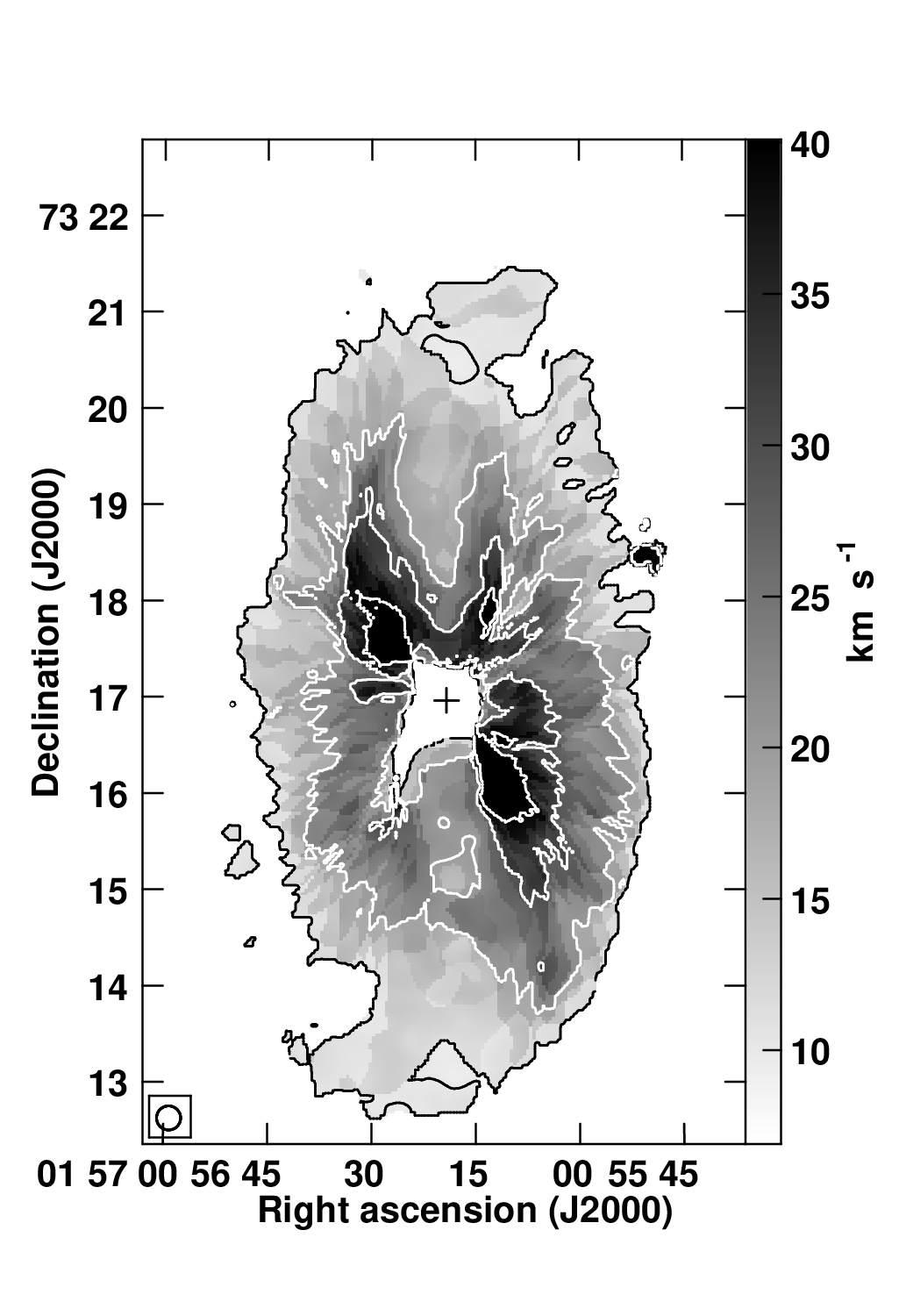}}
\caption{{\bf UGC~1378}: the cross marks the optical centre of the galaxy.
(a) moment 0 map showing H{\sc i} column density in contours  at resolution 
$25^{\prime\prime}\times25^{\prime\prime}$, overlaid on NIR grey sale. 
The contour levels are  1.4, 2.8, 5.6, 11.2, 15.0 $\times$
10$^{20}$ atoms cm$^{-2}$.
(b) moment 0 map showing  H{\sc i} column density in contours at 
resolution $50^{\prime\prime}\times 50 ^{\prime\prime}$, are superposed on DSS B-band
image in grey scale. The contour levels are 0.3, 0.7, 1.4, 2.8, 4.7, 6.0, 7.0
$\times$ 10$^{20}$ atoms cm$^{-2}$.
(c) High resolution H{\sc i} column density contours 
at resolution $15^{\prime\prime}\times15^{\prime\prime}$,
overlaid on NUV grey scale. The contour levels are  4, 7.8, 11.7, 19.6, 32.6  
$\times$ 10$^{20}$ atoms cm$^{-2}$.
(d)  High resolution moment 0 map in  contours and grey scale.  Contour 
levels are similar to Figure \ref{fig:figure.2}(c).
(e) H{\sc i} velocity map (contours and grey scale)  at resolution
$25^{\prime\prime}\times 25^{\prime\prime}$.  
The contours are plotted at  2750, 2800, 2850, 2900, 2950, 
3000, 3050, 3100, 3150 and 3200 km s$^{-1}$ (from north to south).
(f) H{\sc i} moment 2 map (grey scale and contours)  at resolution
$15^{\prime\prime}\times 15 ^{\prime\prime}$.  
The contours are plotted at 10, 20, 30, 40 km s$^{-1}$.}
\label{fig:figure.2}
\end{figure*}

In  Figure \ref{fig:figure.1}(a) the H{\sc i} spectrum produced from 
the spectral cube at $\sim$ $50^{\prime\prime}\times50^{\prime\prime}$ resolution 
is shown. The total H{\sc i} flux is estimated to be $\sim$ 34.5$\pm$2.2 Jy km s$^{-1}$ 
which gives a H{\sc i} mass estimate of 1.2$\pm$0.2  $\times$ 10$^{10}$ M$_{\odot}$ 
for a distance of 38.8 Mpc.

A H{\sc i} hole associated with the optical centre is detected in moment 0 maps.  
A region of about 30$^{\prime\prime}$ ($\sim$ 6kpc) appears to be evacuated 
of H{\sc i}, as seen in the highest resolution map (Figure \ref{fig:figure.2}d).  
This could be a  combination of absorption against the active nucleus 
and lower H{\sc i} surface densities  in the central bar. In our radio 
continuum map at 1.4 GHz \citep{mishra15} a couple of compact sources are 
detected in the central 15$^{\prime\prime}$ region. The observed H{\sc i} 
hole extends over both the sources and continues to the south-east.

The NIR disk of the galaxy is about 2$^{\prime}$ in size 
(Figure \ref{fig:figure.2}(a)) whereas, the H{\sc i} disk 
is about 8$^{\prime}$ in size (about 96 kpc). The morphology 
of the NIR emission with hints of spiral arms arising in the 
north-east and south-west matches with the high resolution H{\sc i} moment-0 
map  high H{\sc i} column density regions are seen
in the north-east and south-west of the centre (Figure \ref{fig:figure.2}(d)). 

The velocity field is regular across the galaxy (Figure \ref{fig:figure.2}(e)).
The dispersion in the outer parts of the disk is 10 km s$^{-1}$ 
increasing to 20 km s$^{-1}$ in the inner 5$^{\prime}$ 
(Figure \ref{fig:figure.2}(f)). The dispersion keeps increasing 
towards the centre of the galaxy and is 40 km s$^{-1}$ in the central
parts especially in the regions on either side of the centre along a position
angle of about 45$^{\circ}$ which coincides with the high column density H{\sc i}. 
This high dispersion region coincides with the NIR disk. 
The GALEX image available for this galaxy is not deep enough and does not 
detect any NUV emission from the galaxy (Figure \ref{fig:figure.2}(c)). 
The radio continuum emission
at 610 MHz is mostly confined to the central parts 
(see Figure \ref{fig:figure.6}(a)) and no H{\sc i} emission 
is seen to be coincident with the radio emission.

The rotation curve of UGC~1378 is shown in Figure \ref{fig:figure.7}(a). 
The rotation curve is flat out to a radial distance of 40 kpc.

\subsection{UGC~1922}

The global  H{\sc i} emission profile of UGC~1922 generated from 
the data at resolution $45^{\prime\prime}\times 45^{\prime\prime}$ 
is shown in Figure \ref{fig:figure.1}(b). The integrated flux is 
$\sim$ 6.0$\pm$1.1 Jy km s$^{-1}$  and the H{\sc i} mass is 3.2$\pm$0.4 $\times$ 
10$^{10}$ M$_{\odot}$ for a distance to the galaxy of 150 Mpc
(refer to Table \ref{tab:lsb_hiresult}).

\begin{figure*}
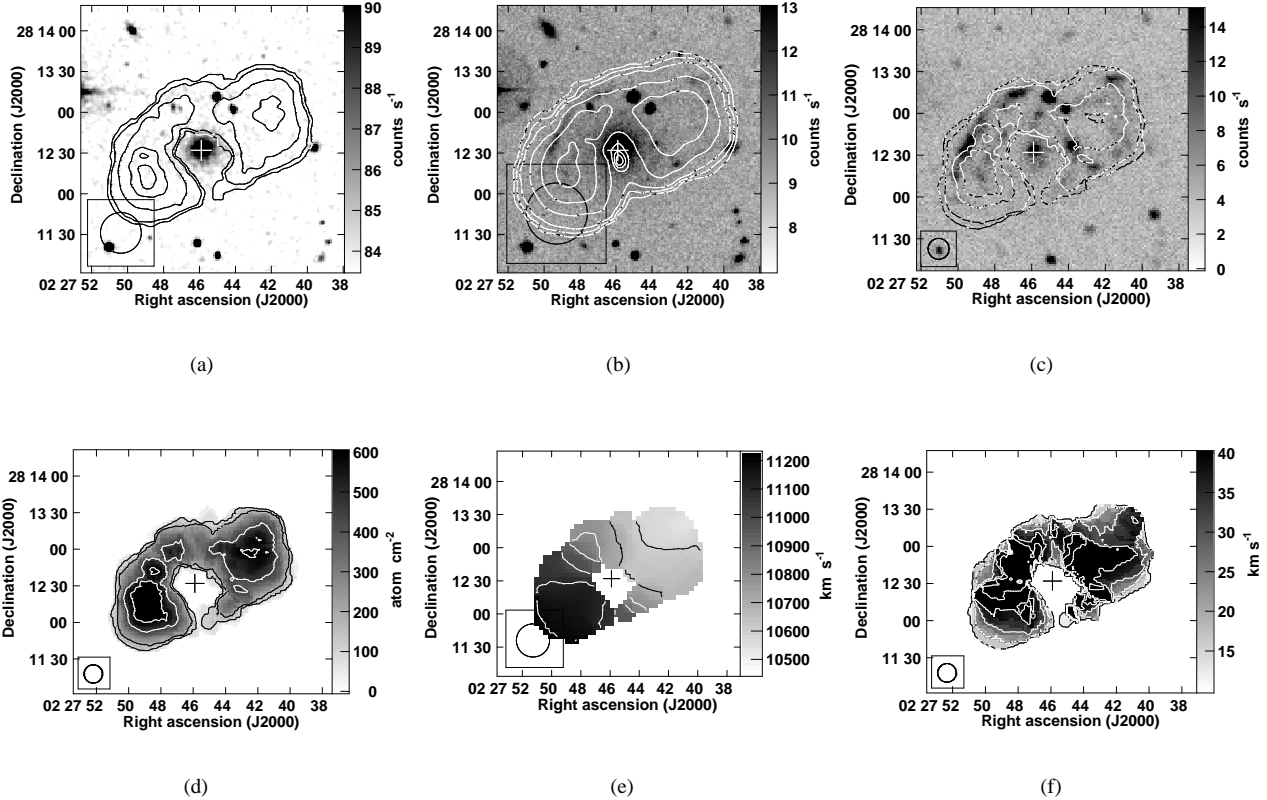

\vspace*{50pt}
\subfigure[]{\includegraphics[height=5.5cm]{figure.3a.ps}}
\subfigure[]{\includegraphics[height=5.5cm]{figure.3b.ps}}
\subfigure[]{\includegraphics[height=5.5cm]{figure.3c.ps}}
\subfigure[]{\includegraphics[height=5.0cm]{figure.3d.ps}}
\subfigure[]{\includegraphics[height=5.0cm]{figure.3e.ps}}
\subfigure[]{\includegraphics[height=5.0cm]{figure.3f.ps}}
\caption{{\bf UGC~1922}: the cross marks the optical centre of the galaxy.
(a) moment 0 map showing H{\sc i} column density in contours at resolution $30^{\prime\prime} 
\times 30^{\prime\prime}$, superposed on NIR grey scale. 
The contour levels are   1, 2, 5, 8 $\times$10$^{20}$ atoms cm$^{-2}$.
(b) Low resolution moment 0 map showing  H{\sc i} column density in contours 
at resolution $45^{\prime\prime}\times 45^{\prime\prime}$, 
superposed on DSS grey scale.
The contour levels are 0.4, 0.8, 1.4, 3, 4.3, 5.8
$\times$10$^{20}$ atoms cm$^{-2}$.
(c) moment 0 map showing H{\sc i} column density at resolution 
$15^{\prime\prime}\times 15^{\prime\prime}$, 
overlaid on NUV grey scale. The contour levels are  4, 8, 19, 26
$\times$ 10$^{20}$ atoms cm$^{-2}$.
(d)  High resolution moment 0 map in  contours and grey scale.  Contour
levels are similar to Figure \ref{fig:figure.3}(c).
(e) Isovelocity H{\sc i} map (contours and grey scale)  at resolution
$45^{\prime\prime}\times 45^{\prime\prime}$.  
The contours are plotted between 10600 (north-west) and 
11200 (south-east) km s$^{-1}$ in steps of 100 km s$^{-1}$.
(f) H{\sc i} moment 2 map (grey scale and contours)  at resolution
$15^{\prime\prime}\times 15 ^{\prime\prime}$.  
The contours are plotted at 10, 20, 30, 40, 50  km s$^{-1}$.}
\label{fig:figure.3}
\end{figure*}

The galaxy shows a central H{\sc i} hole (Figures \ref{fig:figure.3}(a)-(d)) 
likely due to H{\sc i} absorption against the active nucleus.  
Enhanced NUV emission is seen along a broken ring to the north of the optical 
centre of the galaxy coinciding with the H{\sc i} emission. 
The blobby NUV emission skirts the highest H{\sc i} column densities and appears to trace 
a ring around the centre of the galaxy.  
This might indicate that the star formation has been triggered at the edges of the
detected H{\sc i} distribution or has used up high column density H{\sc i} 
coincident with them. The radio continuum emission
at 610 MHz is resolved (see Figure \ref{fig:figure.6}(b)), 
and confined to the central regions of this galaxy \citep{mishra15}.

The velocity field appears to be fairly regular with the gas approaching in the north 
and receding in the southern part  (Figure \ref{fig:figure.3}(e)). The velocity dispersion  
in the central parts is about 50 km s$^{-1}$ as seen in the high resolution maps. 
These motions are likely due to the central AGN.
The rest of the galaxy shows an almost uniform dispersion of 30 km s$^{-1}$
(Figure \ref{fig:figure.3}(f)).  
Some H{\sc i} is also detected from the north of the centre of the galaxy - 
the dispersion of gas here is similar to rest of the galaxy.

The observed rotation curve is shown in  Figure \ref{fig:figure.7}(b).
The rotation curve appears to be rising to the last measured point at a 
radial distance of 50 kpc from the centre of the galaxy.

\subsection{UGC~4422}

The integrated  H{\sc i} spectrum of UGC~4422 generated 
from the data at a resolution $45^{\prime\prime}\times45^{\prime\prime}$
is shown in Figure \ref{fig:figure.1}(c). 
The estimated total integrated flux and the H{\sc i} mass are respectively,  
16.4$\pm$1.4 Jy km s$^{-1}$ and 1.5$\pm$0.5 $\times$ 10$^{10}$ M$_{\odot}$ 
for a distance to the galaxy of 63.4 Mpc (refer to Table \ref{tab:lsb_hiresult}).

\begin{figure*}
\vspace*{50pt}
\subfigure[]{\includegraphics[height=6.0cm]{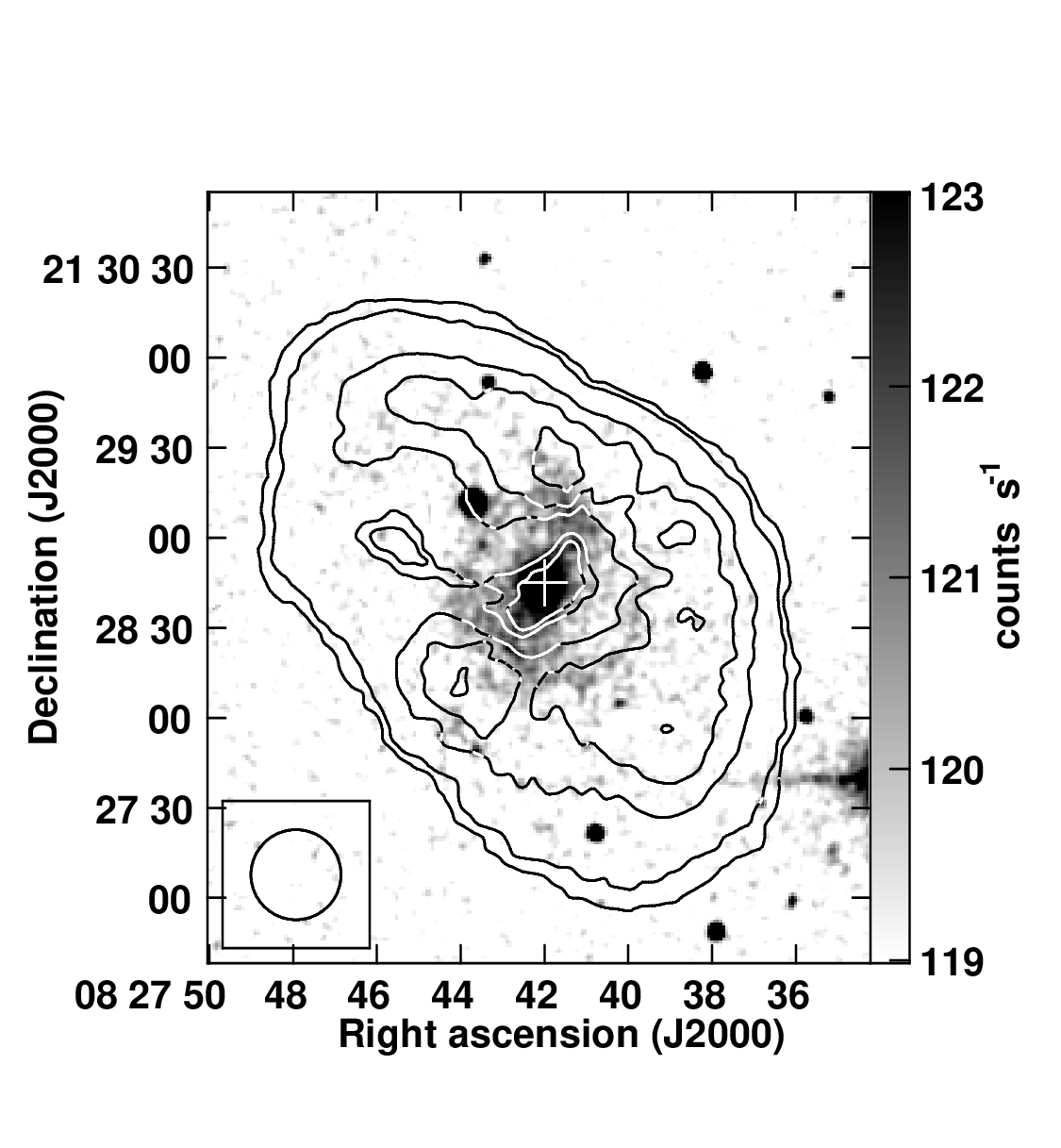}}
\subfigure[]{\includegraphics[height=6.0cm]{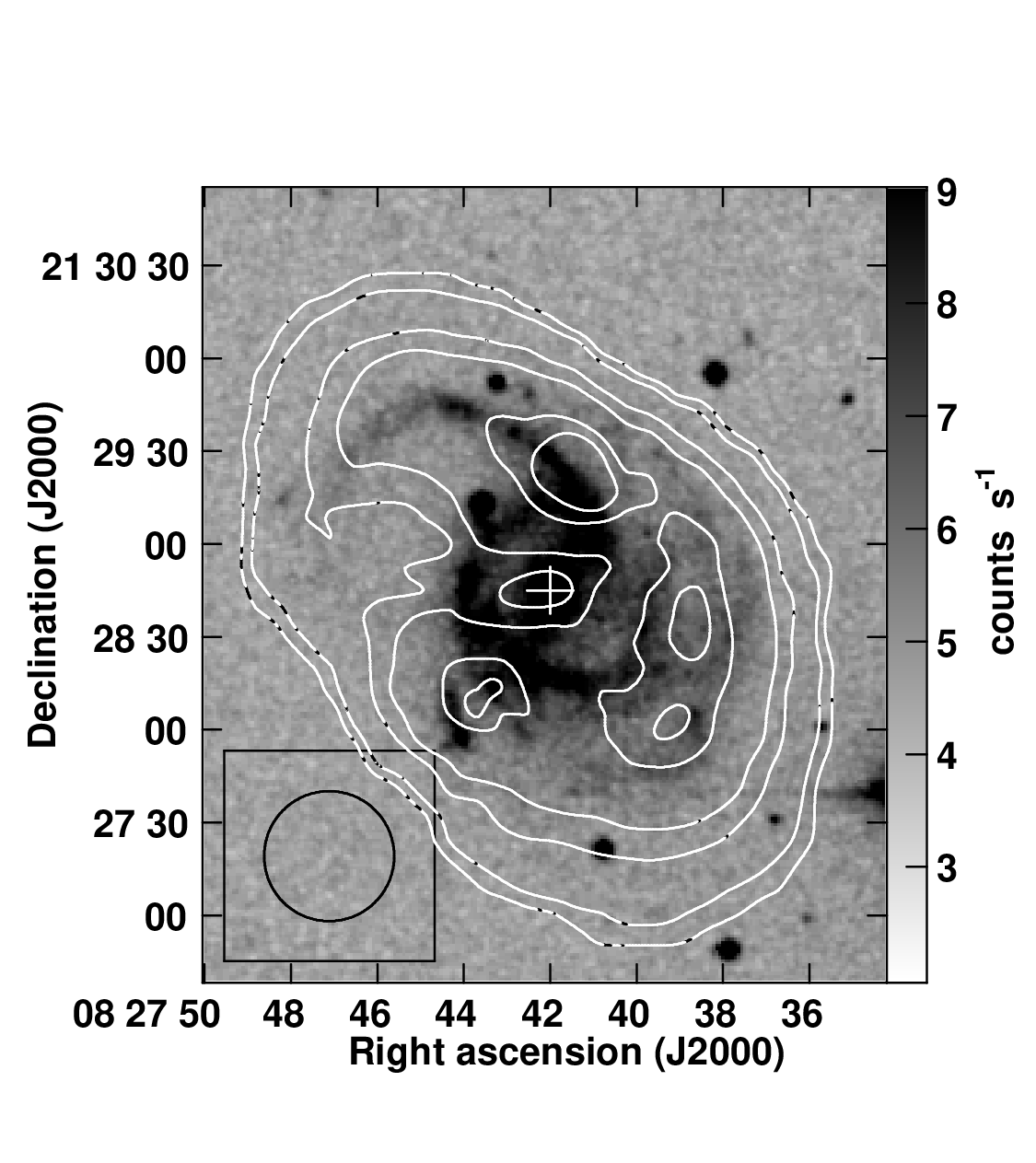}}
\subfigure[]{\includegraphics[height=6.0cm]{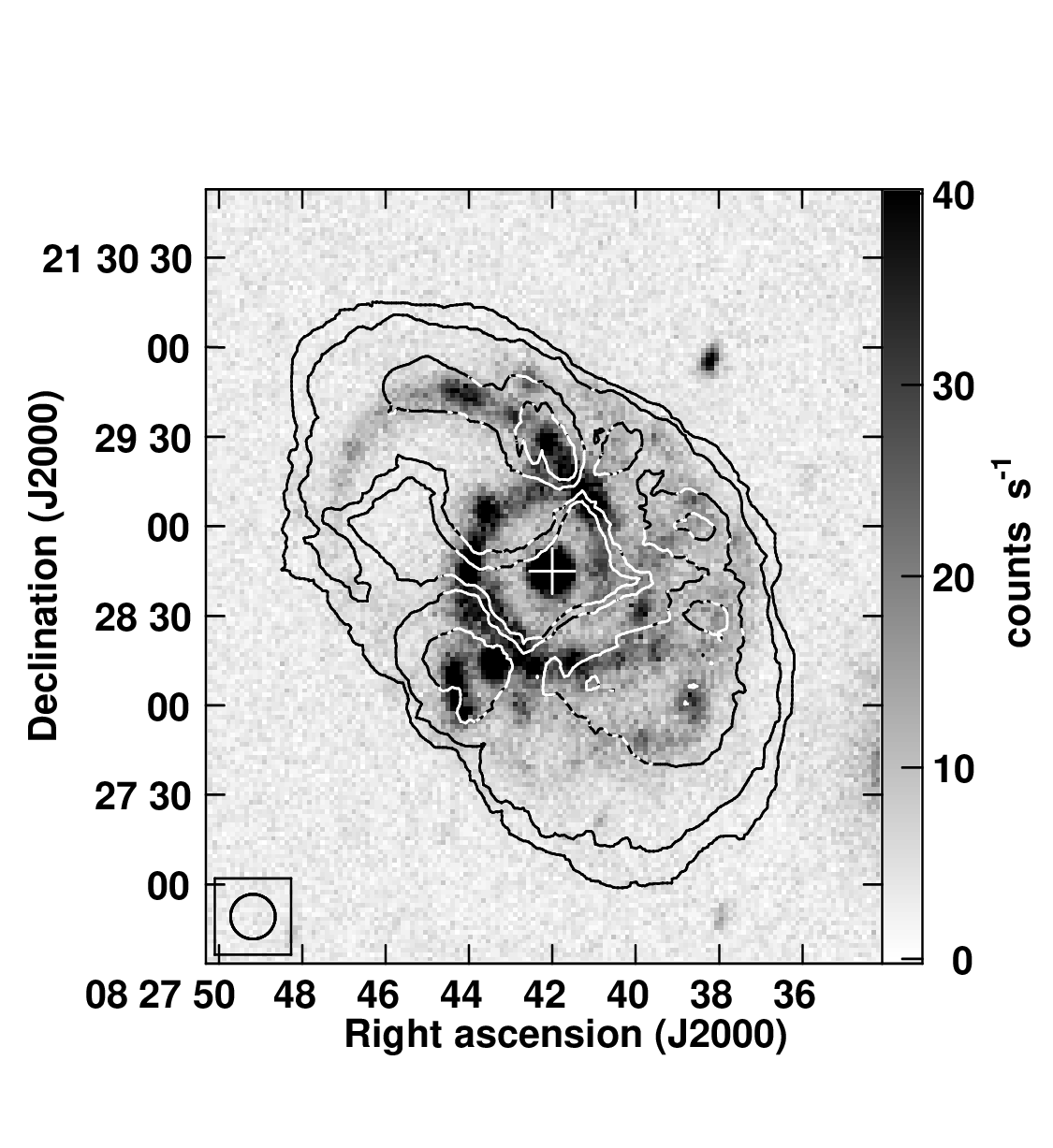}}
\subfigure[]{\includegraphics[height=6.0cm]{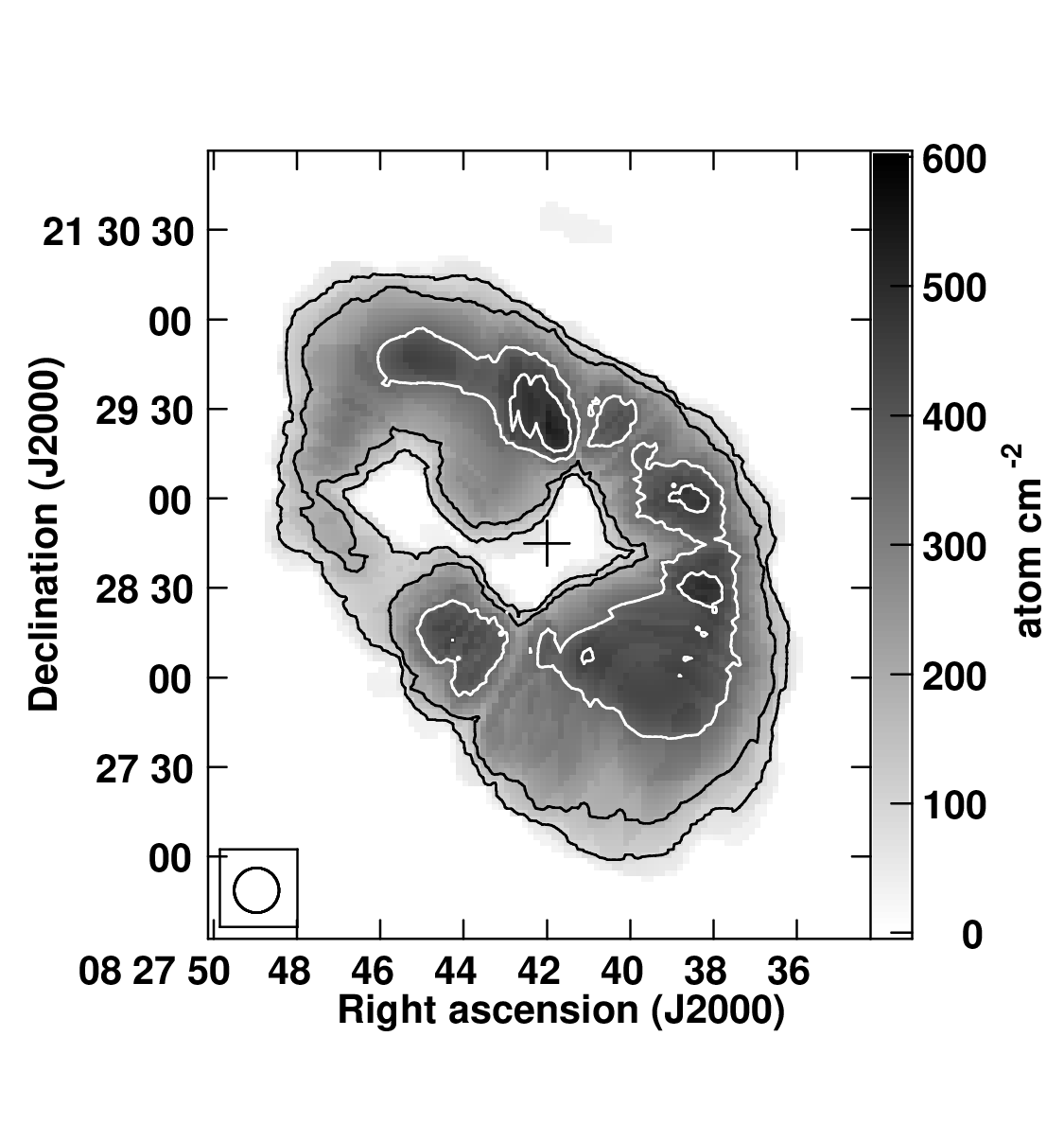}}
\subfigure[]{\includegraphics[height=6.0cm]{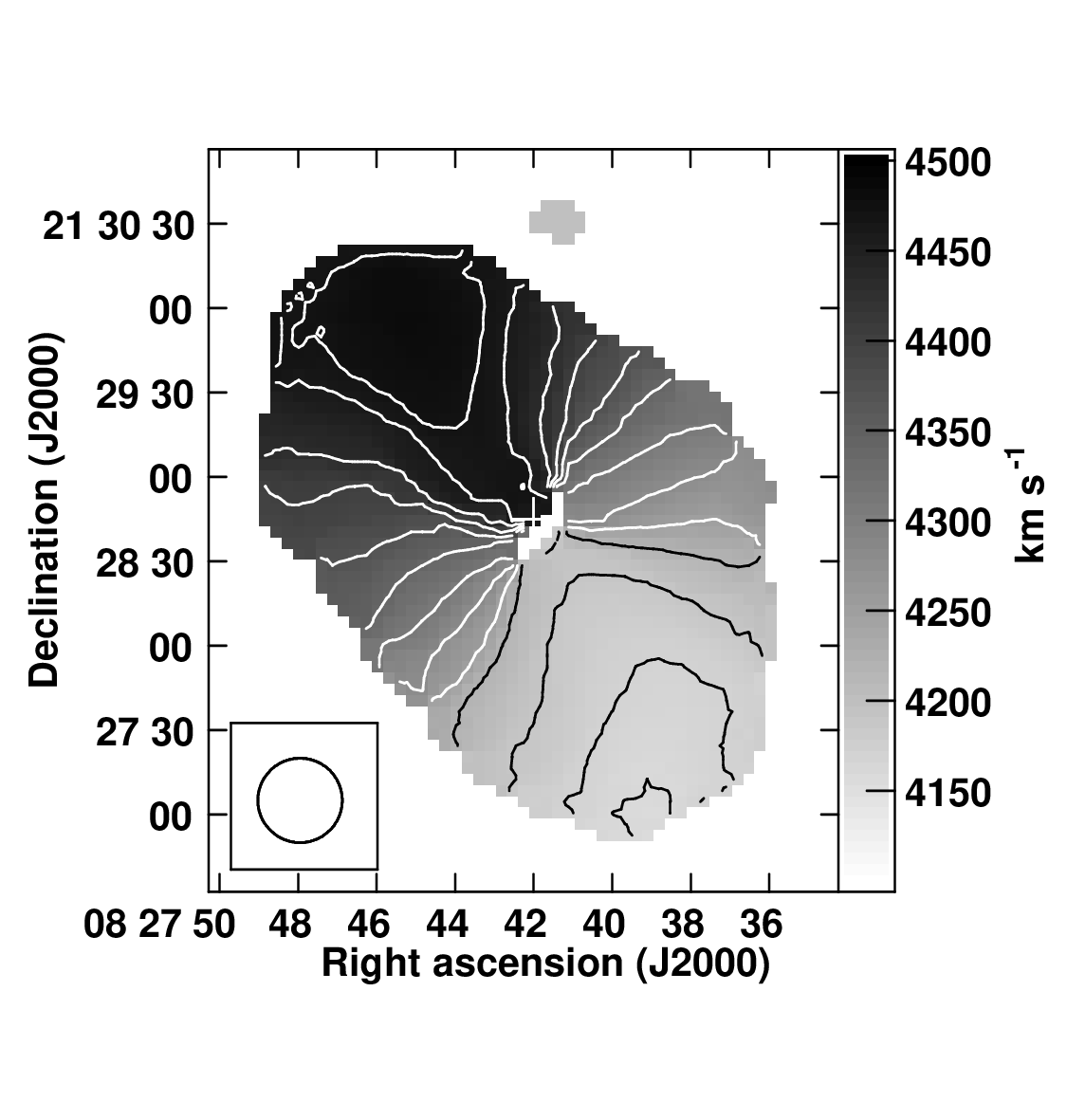}}
\subfigure[]{\includegraphics[height=6.0cm]{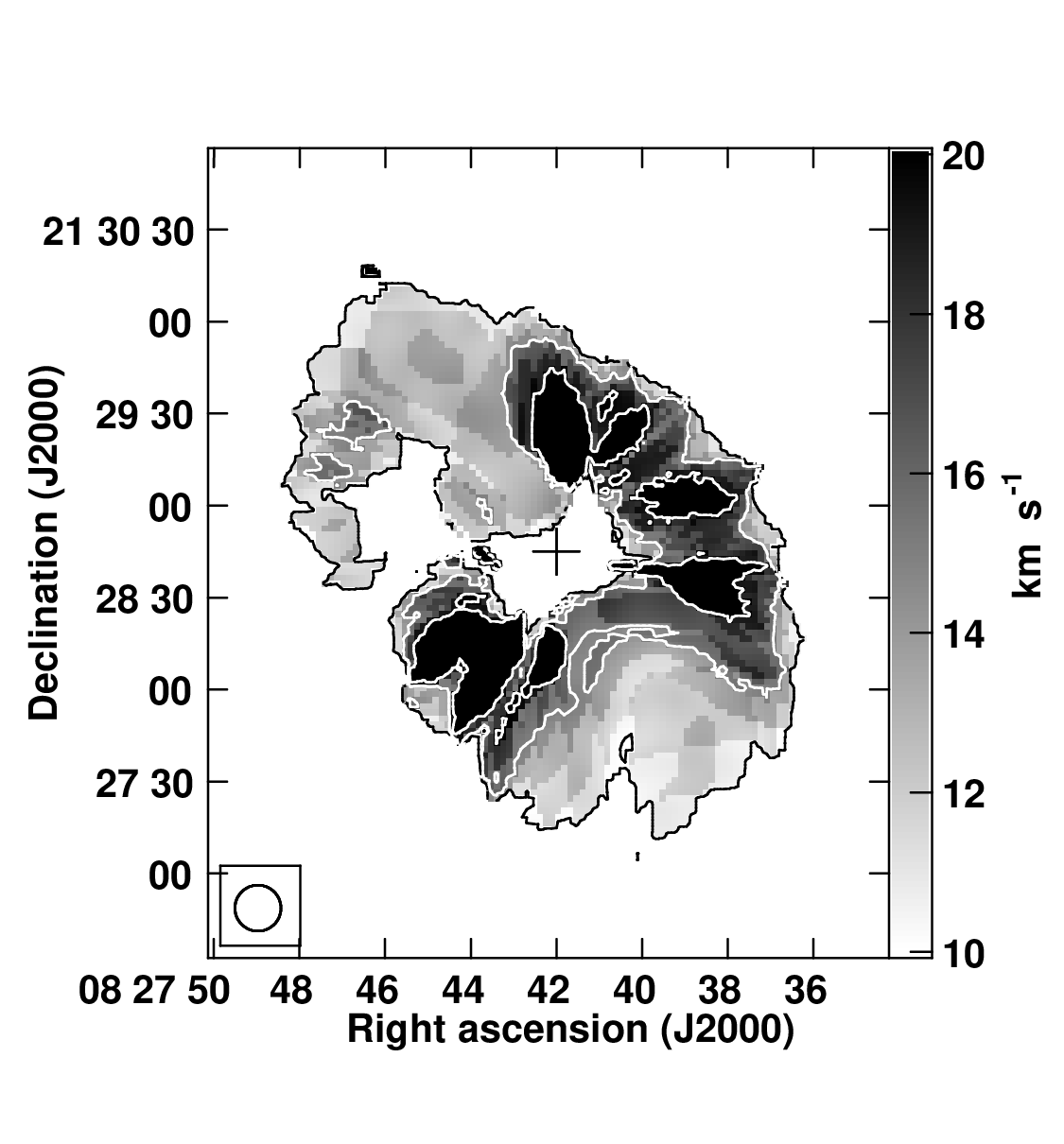}}
\caption{{\bf UGC~4422}: the cross marks the optical centre of the galaxy.
(a) moment 0 map showing H{\sc i} column density in contours 
at resolution $30^{\prime\prime}\times 30^{\prime\prime}$,  
overlaid on NIR  grey scale. The contour levels are 1, 2, 5, 6, 8
$\times$ 10$^{20}$ atoms cm$^{-2}$.
(b) moment 0 map showing  H{\sc i} column density   
in contours at resolution $45^{\prime\prime}\times 45^{\prime\prime}$
overlaid on DSS B-band image in grey scale.  
The contour levels are 0.4, 0.8, 2, 4.3, 4.8 
$\times$ 10$^{20}$ atoms cm$^{-2}$.
(c) H{\sc i} column density in  contours at resolution $15^{\prime\prime} 
\times 15^{\prime\prime}$, superposed on NUV grey scale.
The contour levels are   4, 7.8, 15.6, 19.6 $\times$10$^{20}$ atoms cm$^{-2}$.
(d)  High resolution moment 0 map in  contours and grey scale.  Contour
levels are similar to Figure \ref{fig:figure.4}(c).
(e) Isovelocity H{\sc i} map (contours and grey scale) at resolution 
$30^{\prime\prime}\times 30^{\prime\prime}$. 
The contours are plotted between 4160 (south-west) and 4470 (north-east)  km s$^{-1}$ 
in steps of 30 km s$^{-1}$. 
(f) H{\sc i} moment 2 map (grey scale and contours)  at resolution
$15^{\prime\prime}\times 15^{\prime\prime}$. 
The contours are plotted at 10, 15, 20 km s$^{-1}$.}
\label{fig:figure.4}
\end{figure*}

The H{\sc i} disk is larger than the optical disk and the star formation 
follows the H{\sc i} distribution (Figures \ref{fig:figure.4}(a)-(d)). 
A central hole in the H{\sc i} distribution due to the active nucleus 
is also seen in UGC~4422.
NIR emission is confined to the central regions with diffuse
emission surrounding an intense core.  
N$_{HI}$ in the central parts is $\sim$ 10$^{21}$ atoms cm$^{-2}$.
The H{\sc i} morphology is well-correlated with the NUV morphology 
(Figure  \ref{fig:figure.4}(c)). 
Highest column density clumps in the high angular resolution
maps are clearly associated with vigorous star formation seen in the NUV
(Figure \ref{fig:figure.4}(c)).

The velocity field across the galaxy displayed in  Figure  \ref{fig:figure.4}(e) 
($\sim$ $30^{\prime\prime}\times30^{\prime\prime}$)  
appears undisturbed except for a warp in the outer parts.  Some distortion 
in the velocity field north of the centre is also detected. The H{\sc i} 
velocity field is approaching in the south and receding in the northern parts.
The velocity dispersion shows a smooth gradient from the outer parts to the
inner parts.  It is 10 km s$^{-1}$ in the outer parts and rises to 20 km s$^{-1}$
in the central part (Figure \ref{fig:figure.4}(f)). 
The radio continuum emission at 610 MHz is 
(see Figure \ref{fig:figure.6}(c)) is confined to the central 
regions whereas at 325 MHz \citep{mishra15} coincidence between the H{\sc i} and the radio 
continuum emission is observed. 

The observed rotation curve of this galaxy which is flat to the last
measured point of about 28 kpc is shown in Figure \ref{fig:figure.7}(c).

\subsection{UM~163}

The integrated H{\sc i} spectrum of UM~163 obtained from the cube of  
angular resolution of $45^{\prime\prime}\times45^{\prime\prime}$ 
is shown in Figure \ref{fig:figure.1}(d). The total H{\sc i} flux estimated 
from the spectrum is 0.8$\pm$0.1 Jy km s$^{-1}$, in disagreement 
with the value 4.4 Jy km s$^{-1}$ estimated using single dish 
observations from the Parkes telescope \citep{doyle05}. 
Since their beam size is $\sim$ 14$^{\prime}$.3, 
the observations are possibly picking up extended emission that GMRT observations are 
not sensitive to.  The H{\sc i} mass corresponding to our observed  flux is 
0.34$\pm$0.04 $\times$ 10$^{10}$ M$_{\odot}$ for the galaxy distance of 136 Mpc
(refer to Table \ref{tab:lsb_hiresult}).  However using the \cite{doyle05} result,
the H{\sc i} mass associated with the galaxy could be higher by at least five times.

\begin{figure*}
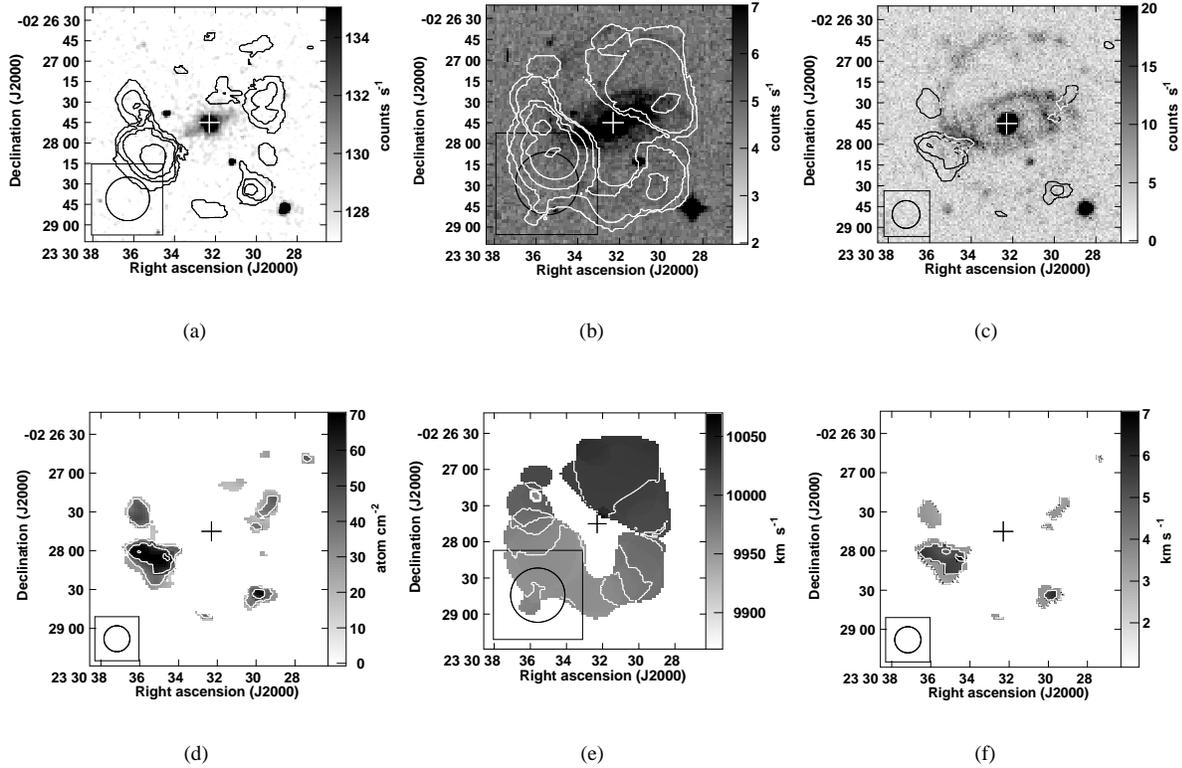

\vspace*{50pt}
\subfigure[]{\includegraphics[height=5cm]{figure.5a.ps}}
\subfigure[]{\includegraphics[height=5cm]{figure.5b.ps}}
\subfigure[]{\includegraphics[height=5cm]{figure.5c.ps}}
\subfigure[]{\includegraphics[height=5cm]{figure.5d.ps}}
\subfigure[]{\includegraphics[height=5cm]{figure.5e.ps}}
\subfigure[]{\includegraphics[height=5cm]{figure.5f.ps}}
\caption{{\bf UM 163}: the cross marks the optical centre of the galaxy.
(a) moment 0 map showing H{\sc i} column density  in contours at a 
resolution of $32^{\prime\prime}\times 32^{\prime\prime}$,  
overlaid upon NIR grey-scale. The contour levels are at H{\sc i} column density
of 0.3, 0.6, 1.0, 1.2 $\times$  10$^{20}$ atoms cm$^{-2}$.
(b) moment 0  H{\sc i} column density  at resolution 
$45^{\prime\prime}\times 45^{\prime\prime}$,  
overlaid on DSS B-band image in grey scale.
The contour levels are  0.1, 0.3, 0.6, 
0.8, 1.0 $\times$ 10$^{20}$ atoms cm$^{-2}$.
(c) H{\sc i} column density contours at resolution of $20^{\prime\prime}
\times 20^{\prime\prime}$, overlaid upon NUV grey-scale.
The contour levels are  0.8, 1.3, 1.8 $\times$  10$^{20}$
atoms cm$^{-2}$.
(d)  High resolution moment 0 map in contours and grey scale.  Contour
levels are similar to Figure \ref{fig:figure.5}(c).
(e) Velocity  field (contours and grey scale) at resolution 
$45^{\prime\prime}\times 45^{\prime\prime}$.
The contours are plotted between 9950 (south-east) and 10020 
(north-west) km s$^{-1}$ in steps of 10 km s$^{-1}$.
(f) H{\sc i} moment 2 map (grey scale and contours)  at resolution
$20^{\prime\prime}\times 20^{\prime\prime}$.  
The contours are plotted at 3, 5, 7 km s$^{-1}$.}
\label{fig:figure.5}
\end{figure*}

The H{\sc i} from this galaxy is detected along a disk which 
is broken in the north (Figures \ref{fig:figure.5}(a)-(d)). 
No H{\sc i} is detected from the optical centre of the  galaxy. 
Highest  H{\sc i} column 
densities are observed towards the eastern part of the galaxy. 
The peak column density that we observe from this galaxy is about
$2.1\times10^{19}$ atoms cm$^{-2}$. 

The DSS B band image detects a large bar-like structure extending out
from the central source (Figure \ref{fig:figure.5}(b)). 
While higher H{\sc i} column densities are detected at the edges of
the bar, no H{\sc i} is detected from the bar. It is possible that the 
H{\sc i} gas has been driven into the corotational ring by the strong bar. 
This may account for the lower H{\sc i} surface densities over the entire 
bar region. The SDSS optical images show a distinct circum-nuclear ring 
in the center, which could be due to nuclear star formation again 
triggered by the bar. Such behaviour has been observed
in other barred spirals also.
The NUV detects a ring around the central source with it emerging from 
the east and continuing north. The H{\sc i} is confined to a disk in the south 
with little correlation seen with the NUV (Figure \ref{fig:figure.5}(c)) along it. 
 
The velocity field (Figure \ref{fig:figure.5}(e)) 
smoothly varies along the disk from 9950 km s$^{-1}$ in the east to 10020
km s$^{-1}$  in the north-west
region.  However, higher velocities are seen at the tip of the disk in the east 
and  both ends of the broken disk are moving at higher velocities compared 
to the central region. 
The velocity dispersion along the disk is low; ranging from 3 to 7 km s$^{-1}$ 
(Figure \ref{fig:figure.5}(f)). 

\begin{figure*}
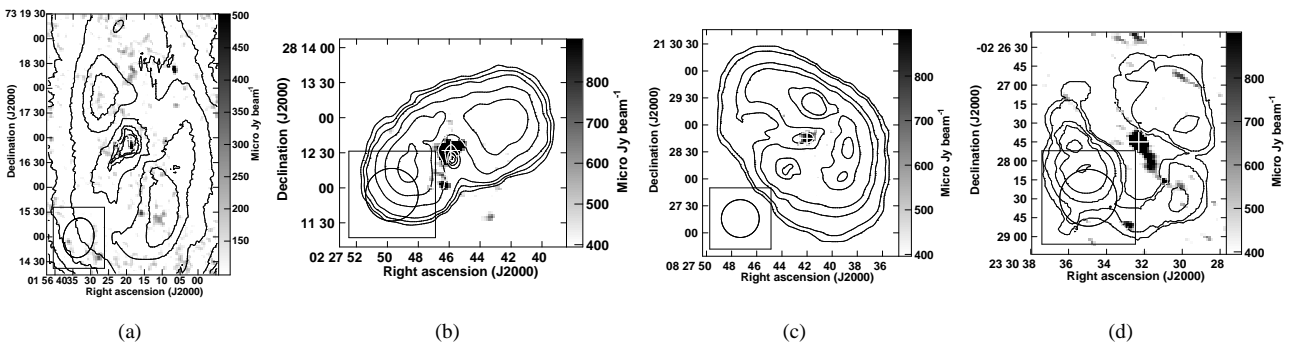

\vspace*{50pt}
\subfigure[]{\includegraphics[height=4.5cm]{figure.6a.ps}}
\subfigure[]{\includegraphics[height=4.5cm]{figure.6b.ps}}
\subfigure[]{\includegraphics[height=4.5cm]{figure.6c.ps}}
\subfigure[]{\includegraphics[height=4.5cm]{figure.6d.ps}}
\caption{Zoomed-in  MOM0  contours  are  overlaid on 610 MHz map (grey scale).
{\bf (a)} UGC~1378  {\bf (b)} UGC~1922  {\bf (c)}  UGC~4422  {\bf (d)}  UM~163}
\label{fig:figure.6}
\end{figure*}

A likely explanation for the  low H{\sc i} surface densities in the northern 
part of the disk of the galaxy could be that the recent burst of star formation 
has exhausted the high column density gas. Only low column density gas exists 
which our observations might not be sensitive to i. e. column densities lower 
than 1$\times$ 10$^{19}$ atoms cm$^{-2}$. This needs to be further investigated 
and understood.  The radio continuum emission at 610 MHz is confined to the central
regions (see Figure \ref{fig:figure.6}(d)). No correlation is noticed 
 between
H{\sc i} and radio continuum emission. 

\section{Discussion}

\subsection{Rotation Curves of GLSB galaxies}

We have estimated the rotation curves using the AIPS task GAL on the 
velocity field of the galaxies and determined their H{\sc i} centres, 
inclination, position angle and rotation velocity. 
The rotation curves are obtained using the tilted ring model \citep{begeman89}. 
The rotation curve fitting procedure involves using the observed velocity
field to derive five galaxy properties namely dynamical centre (X,Y), 
systemic velocity (V$_{sys}$), position angle (PA), inclination (Inc) and rotation 
velocity (V$_{rot}$). The velocities were averaged in elliptical annuli of width 10$^{\prime\prime}$ .
The centre, PA and the Inc estimated from H{\sc i} moment 0 map were taken as initial 
guesses. In the first iteration we kept all the parameters free. The inclination and
centre was determined for each annulus and found to be more or less similar for all annuli 
for UGC~1378 and UGC~4422. In the next iteration, the inclination was fixed to the 
average value from the first iteration. The resulting rotation curves for UGC~1378 and 
UGC~4422 from this exercise are shown in Figure \ref{fig:figure.7}(a) and (c). 
The estimated rotation velocities are 282 and 254 km s$^{-1}$. We could not fit 
a rotation curve to the velocity field of UM 163.  
The rotation curves were derived using the mid-resolution 
$\sim$ ($25^{\prime\prime}\times 25^{\prime\prime}$) velocity fields.  
The details  of the fit are listed in Table \ref{tab:lsb_hiresult}. 

The fitting for UGC~1922 was more involved.  When all the parameters were kept 
free then we  found that the inclination and PA showed variations with radial 
distance from the centre of the galaxy and the resulting rotation curve is 
shown by stars in  Figure \ref{fig:figure.7}(b). The estimated rotation velocity 
is very large at $\sim$ 658$\pm$42. From this run of GAL, we estimated an 
average value for the inclination which was  32 degrees and  PA of 132 
degrees which were fixed in the next iteration of GAL. This resulted in 
a rotation curve shown by the filled circles in  Figure \ref{fig:figure.7}(b) 
and a rotation velocity of 588$\pm$32 km s$^{-1}$. This is also 
large and not commonly observed. To further check this, we derived the 
inclination from the observed H{\sc i} ellipticity which was 51$\pm$2 degrees.  
We re-ran GAL by fixing the inclination to 51 degrees and the resulting 
rotation curve gave a rotation velocity of 432$\pm$12 km s$^{-1}$ as shown 
by the filled squares in Figure \ref{fig:figure.7}(b).  Although this is 
also large, we believe that till further observations can throw more 
light on this intriguing galaxy, this is the conservative estimate of the 
rotation velocity of UGC~1922. We believe that the rotation curve of UGC~1922 
can be confirmed or otherwise if a better estimate of the inclination of 
the galaxy can be obtained. As noted above, if the inclination 
of UGC~1922 is 32 degrees then its rotation velocity comes out to be 588 km s$^{-1}$ 
and if it is 51 degrees it is 432 km s$^{-1}$.  Nevertheless, UGC~1922 appears 
to be among the fastest rotating galaxies, which to the best of our knowledge are 
UGC 12591 (506 km s$^{-1}$; \citealt{giovanelli86}) and NGC 1961 (402 km s$^{-1}$; 
\citealt{courtois15}; \citealt{haan08}) as reported earlier in the literature. 
It is obvious that such galaxies are rare.

The position angle in all galaxies showed a slow variation of about $5^{\circ}$. 
The rotation curves do not show any drop at large radial distance from
the centre.  Similarly flat rotation curves at large radial
distances are also noted in UGC~2936 \citep{pickering99},  
UGC~6614 and Malin~2 \citep{pickering97}.
Thus all the six GLSB galaxies out of the total seven studied in radio continuum by
\cite{mishra15} show a flat rotation curve to the outermost observed point  -
i.e. to a radial extent of about 42 kpc (220$^{\prime\prime}$), 
51 kpc (70$^{\prime\prime}$) and 28 kpc (90$^{\prime\prime}$) 
in UGC~1378, UGC~1922 and UGC~4422 whereas the radial extent of the 
optical disk is 20 kpc,  44 kpc and 15 kpc respectively.  
Similarly a flat rotation curve is observed upto radial distances of 35 kpc in 
UGC 2936 \citep{pickering99}, 50 kpc in UGC~6614 and 80 kpc in Malin~2 
\citep{pickering97} whereas the optical radial extent for these galaxies are
19 kpc, 21 kpc and 41 kpc.  Thus, the H{\sc i} disk is roughly twice the size
of the optical disk for all the GLSB galaxies and the rotation curve
is flat over the detectable H{\sc i} disk.  Flat rotation curves have also
been observed in LSB spirals - five LSB spirals in 
\cite{vanderhulst93},  19 LSB spirals in \cite{deblok96a} although we 
note that one of the curves appears to show a drop and eight 
LSB galaxies in \cite{pizzella05}.  

The rotation velocities of the GLSB galaxies range from  225 to 432 km s$^{-1}$. 
The dynamical mass estimated for the three galaxies observed here ranges 
from few times $10^{11}$ to few times $10^{12}$ M$_\odot$.  
Thus except for UM~163 where H{\sc i} is detected in a disk around the central bar of 
the galaxy and the line width is only about 110 km s$^{-1}$, all the GLSB 
galaxies are fast rotators.  However we note that the LSB galaxies have lower
rotation velocities between 30 and 120 km s$^{-1}$ \citep{deblok96a}.

LSB galaxies seldom seem to show a drop in the rotation velocity in the 
outer regions as seen in the sample of six GLSB galaxies; including the sample 
of Pickering et al. (1997, 1999) and 24 LSB galaxies (\citealt{vanderhulst93}; 
\citealt{deblok96b}).  One of the galaxies from the \cite{deblok96b} seems 
to suggest a slight fall in the rotation velocities in the outer parts. 
A few HSB galaxies have been found to show a drop in the rotation
velocities in the outer parts of the galaxies \citep[e.g][]{honma97}.
Whether this difference is significant needs to be investigated 
using a larger sample of LSB galaxies.

\subsection{H{\sc i} properties}

The peak H{\sc i} surface density estimated from the maps made with a beamsize between
25$^{\prime \prime}$ and 30$^{\prime \prime}$ is lowest at $3.3\times10^{19}$ 
cm$^{-2}$ for UM~163 and highest at $3.2\times10^{21}$ cm$^{-2}$  
for UGC~1378.   For the maps with a resolution close to 15$^{\prime \prime}$, 
we detect gas with peak surface densities of $7.5\times10^{21}$ cm$^{-2}$ 
in the centre of UGC~1378.   Thus, such high surface density H{\sc i} clumps 
are likely to be present in LSB galaxies explaining the observed star formation.  
The peak surface densities detected in the six galaxies observed 
by \cite{vanderhulst93}  at a resolution of about 25$^{\prime \prime}$
was between $5-10 \times 10^{20}$ cm$^{-2}$.   For a resolution of about 
22$^{\prime \prime}$, \cite{pickering99} recorded a peak surface density
ranging from  $4.7-8.5 \times 10^{20}$ cm$^{-2}$ for the three 
GLSB galaxies that they studied.   

All the four galaxies studied here show a hole in the centre of the 
H{\sc i} distribution where an active nucleus is present.  The H{\sc i} is 
likely absorbed against this.
The H{\sc i} appears to avoid the central bar as seen in both 
UGC~1378 and UM~163.  This has been noted in HSB galaxies also.  
All the galaxies show some lopsidedness as seen in their integrated spectra.   
All the GLSB galaxies have large H{\sc i} content.  
The H{\sc i} mass estimated for the seven galaxies lies in the range 
$\sim 0.3-4\times10^{10}$ M$_{\odot}$. The H{\sc i} distribution and star 
formation are well-correlated for two (UGC~1922, UGC~4422) of
the four galaxies studied here whereas no correlation is seen for one of
the galaxies UM~163.

We note that UGC~1922, UGC~4422 and UM~163 have been catalogued as members 
of the Low Density Contrast Extended (LDCE) groups ({\citealt{crook07}; \citealt{crook08}).  
UGC~4422 is also classified as member of a Lyon Galaxy Group 
(LGG; \citealt{garcia93}).  UGC~1378 is not classified as a group member.
It can be assumed that the tidal interactions in the LDCE groups are weaker
than in the LGG groups.  The velocity field of
the four galaxies presented here appear undisturbed.  The velocity field of UGC~2936
\citep{pickering99} also appears undisturbed.  
The kinematics of UGC~6614, which has a low inclination, shows a 
peculiar behaviour with the major axis of
the H{\sc i} distribution being perpendicular to kinematical major axis and could be
a result of a warp \citep{vanderhulst93}.  We could not find information
on the detailed velocity field of Malin~2.  However the undistorted velocity
fields of at least five galaxies support the picture of these galaxies being subjected
to none or only weak tidal forces since they evolve in under-dense 
regions \citep{bothun93}. 

As pointed out by \cite{vanderhulst93} from their study of six LSB
galaxies in H{\sc i}, these galaxies have normal H{\sc i} mass to light
ratios and also the fraction of dynamical mass in H{\sc i};
the differences are that LSB galaxies have more extended H{\sc i} and average
surface densities which are about a factor of 2 lower than in HSB spirals.
LSB disks have a mean younger age compared to HSB disks as deduced
from their bluish colours in the optical \citep{vanderhulst93}.
The H{\sc i} mass of these LSB galaxies ($\sim 10^9$ M$_\odot$; 
\citealt{vanderhulst93}) is an order of magnitude lower than that 
of the seven GLSB galaxies ($\sim 10^{10}$ M$_\odot$)
discussed here.  The fraction of dynamical mass in H{\sc i} is comparable
at $1-4 \%$ in GLSB galaxies and $2-9\%$ for most LSB galaxies 
(\citealt{vanderhulst93}; \citealt{deblok96a}).

LSB galaxies follow the same Tully-Fisher (TF) relation as HSB galaxies 
(\citealt{zwaan95}; \citealt{mcgaugh00}; \citealt{mcgaugh05}). This implies that 
LSB galaxies compared to HSB galaxies of similar properties have
M/L and sizes larger by factor of two.  We find that
out of the sample of seven GLSB galaxies discussed here six 
follow the TF relation. UM~163 is an outlier whose B band magnitude 
is similar to the other LSB galaxies but the line width is narrow.  
This is the galaxy in which the H{\sc i} is observed in a broken disk.

\subsection{Star formation along rings}
The star formation traced by the NUV and FUV emission from GALEX in the 
GLSB galaxies appears to indicate that recent star formation is along rings or 
broken rings.  Several rings of NUV emission are seen in the GLSB galaxies - 
UGC~6614 and Malin~2 (see Figure 6, 7 in \citealt{mishra15}).  
Enhanced NUV emission along broken rings are
visible in UGC~1922 (Figure \ref{fig:figure.3}(c)), UGC~4422 
(Figure \ref{fig:figure.4}(c)) and UM~163 (Figure \ref{fig:figure.5}(c)).   
Out of the total seven GLSB galaxies studied by \cite{mishra15}, five galaxies
show star formation along rings around the centre of the galaxies. 
We also examined a sample of 16 LSB galaxies  whose GALEX results are 
presented in \cite{boissier08} to search for star formation along rings.  
However except for a couple of possible cases,
we did not find this to be a common occurrence in that sample.  This
difference in GLSB galaxies and LSB galaxies could  possibly indicate 
lack of sufficiently dense H{\sc i} gas in the smaller LSB galaxies as 
compared to GLSB galaxies or sensitivity of the NUV observations.  
While such rings have been commonly observed in
both atomic gas and in optical emission in tidally interacting galaxies,
their presence in relatively isolated galaxies argues for a modified explanation. 
One possible explanation we suggest is that a  burst of star formation is
spread along a ring over a rotation period for relatively isolated galaxies. 
The trigger of this star formation could be a weak tidal interaction or a 
random local perturbation in sufficiently dense neutral gas.  

In isolated galaxies, the perturbation would travel along the
ring over a rotation period and if sufficiently dense gas is available
along the ring, can trigger star formation along it.
Three such rings of star formation have been seen in the isolated
spiral NGC~7217 \citep{verdes95}. Since such star forming 
rings are also observed in galaxies
subject to tidal interactions, there should be a way to distinguish between
these two scenarios. If the galaxy was subject to strong tidal
interactions, it can lead to gas condensing to higher densities at several
locations. Subsequently star formation can be simultaneously triggered at
several high density locations in a galaxy including along a ring. In this scenario,
the age of the stars in the entire ring will be similar.
On the other hand, in the isolated galaxy/GLSB scenario, since the differential
rotation is held  responsible for propagating the star formation along a ring,
the stellar ages are likely to show a gradient along the ring
to within a rotation period.   It would be interesting to verify this
by studying the stellar ages along rings in tidally interacting galaxies
and in GLSB galaxies.

\begin{figure}
\subfigure[]{\includegraphics[height=8.5cm,angle=-90]{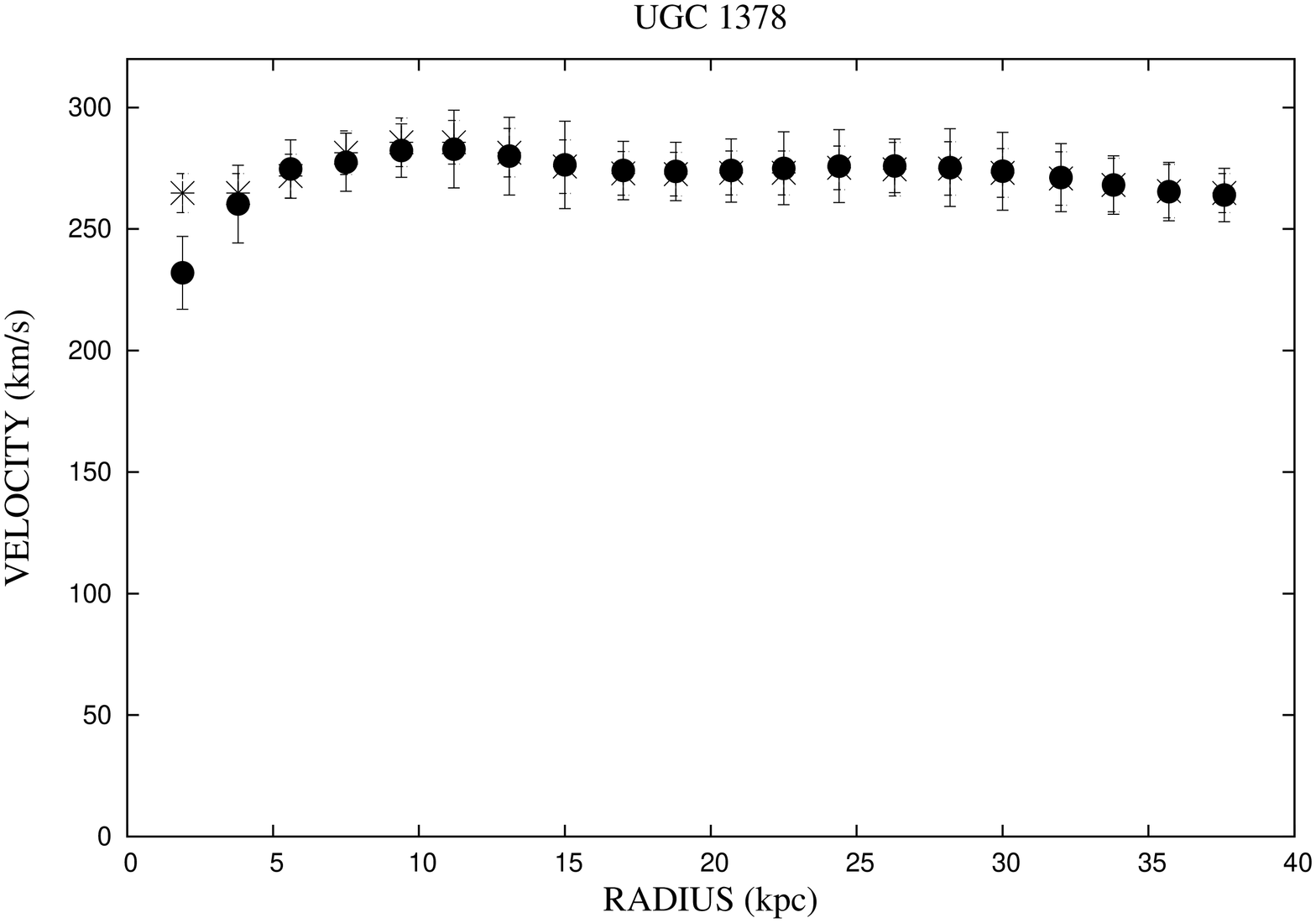}}
\subfigure[]{\includegraphics[height=8.5cm,angle=-90]{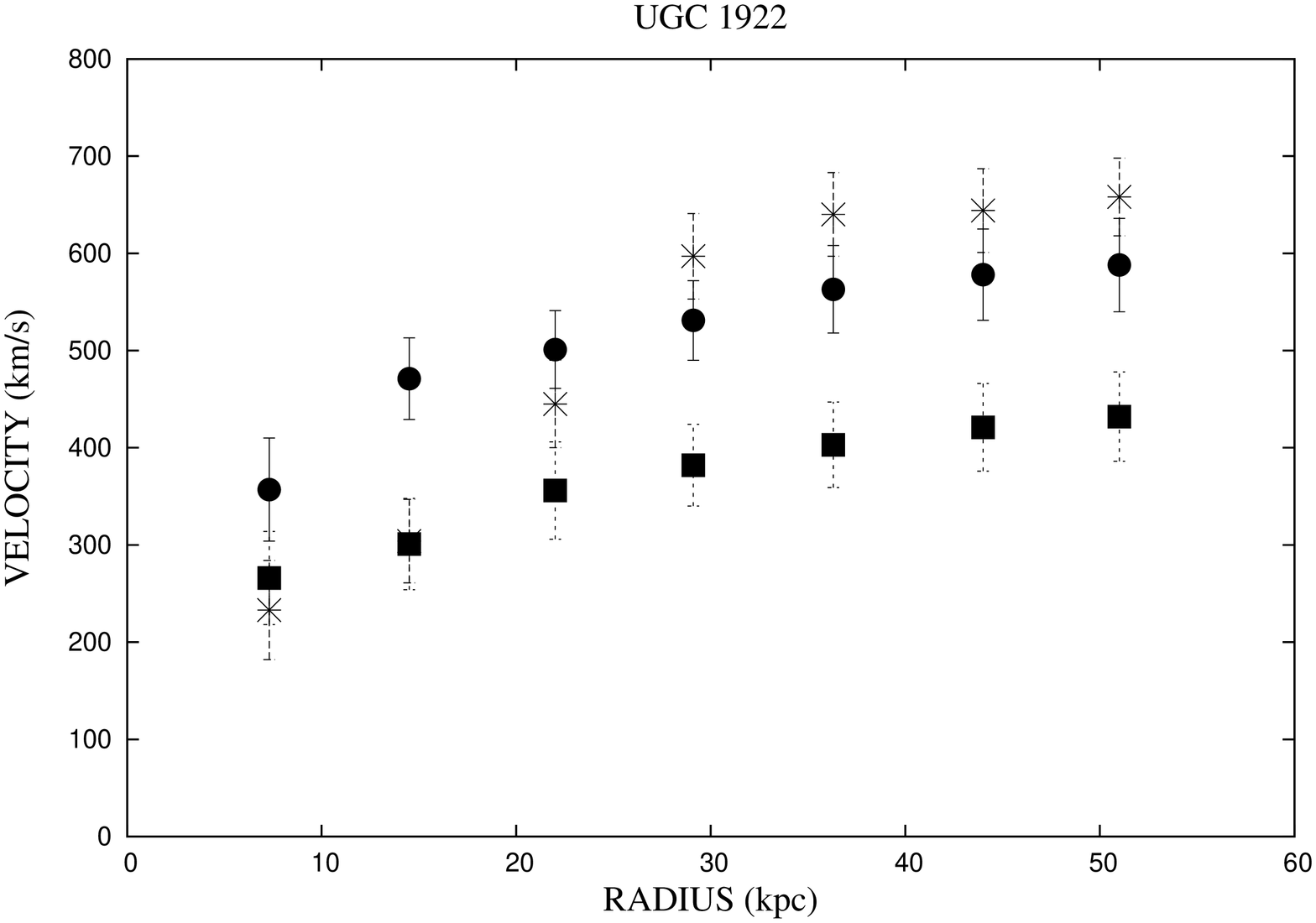}}
\subfigure[]{\includegraphics[height=8.5cm,angle=-90]{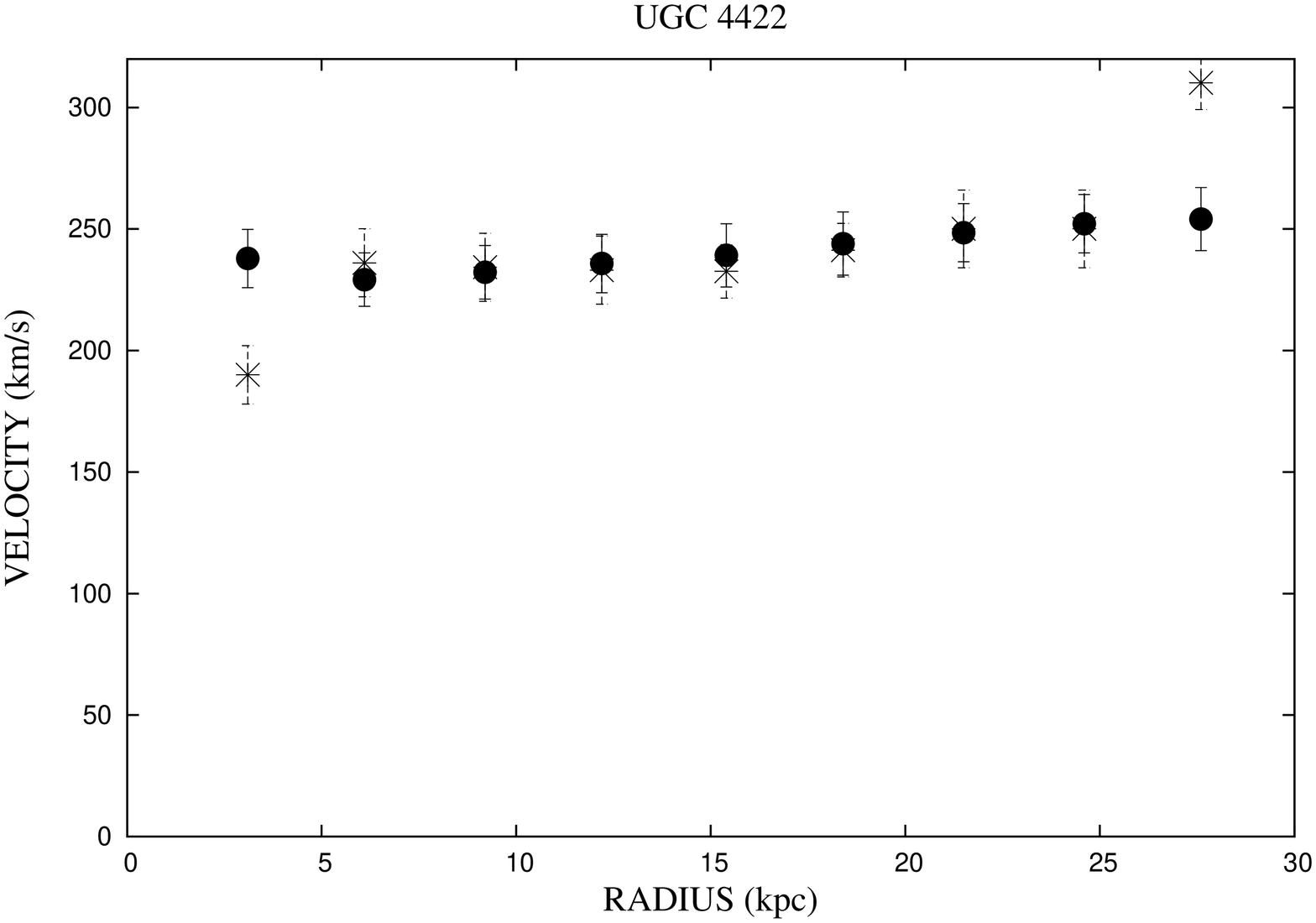}}
\caption{Rotation curves: {\bf (a)} UGC 1378, {\bf (b)} UGC 1922 and {\bf (c)} UGC 4422. 
The filled circles represent variation in V$_{rot}$ with fixed inclination  and centre 
and stars represent variation in V$_{rot}$ with all free parameters. In UGC 1922 filled 
squares show V$_{rot}$ with fixed inclination (derived from  ellipticity of H{\sc i} and centre.}
\label{fig:figure.7}
\end{figure} 

\section{Summary}

In this paper we presented  the H{\sc i} emission study of a sample of
four giant LSB galaxies - UGC~1378, UGC~1922, UGC~4422 and UM~163- using 
GMRT to examine the morphology and dynamics.  We observed extended gas 
disks and found the H{\sc i} masses $\geq$ 10$^{9}$ M$_{\odot}$.
A double-horned H{\sc i} profile similar to HSB  galaxies is observed in 
the sample galaxies.  

The following are the main results from the study:

(1) All the GLSB galaxies have H{\sc i} disk roughly twice the optical sizes. 
The H{\sc i} disks are seen mildly lopsided in the integrated line profiles.  
All the galaxies show a hole in their H{\sc i} distribution in the centres of 
the galaxies likely due to the active nucleus.  

(2) The peak H{\sc i} surface densities are several 
times $10^{21}$ cm$^{-2}$ for most of the GLSB galaxies.
This seems to suggest that there do exist regions of 
high surface densities in GLSB galaxies, sufficient to support star formation.   

(3) The rotation curves of GLSB galaxies and LSB galaxies are flat out to 
the last measured points. This is seen for six (excepting UM 163) of the seven 
GLSB galaxies  and in 24 LSB galaxies (\citealt{vanderhulst93}; \citealt{deblok96a}) 
all derived from H{\sc i} data which measure the rotation beyond the optical disk.  

(4) Six of the seven GLSB galaxies follow the same TF relation 
as HSB galaxies.  UM~163 is an outlier with its blue luminosity 
being similar to the other GLSB galaxies but the H{\sc i} line width being
narrow.

(5) Six of the seven GLSB galaxies show star formation along  
ring(s) or broken ring(s) around the centre of the galaxy as 
inferred from enhanced NUV emission seen in GALEX images.  
It is difficult to comment on the seventh galaxy UGC 1378 
where only a shallow GALEX image exists.  
Since there is a deficit of galaxies within 2 Mpc around LSB galaxies
\citep{bothun93}, the existence of rings cannot be attributed 
to strong tidal interaction. We suggest that in these relatively 
isolated galaxies, star formation is triggered
locally where sufficient H{\sc i} surface densities are available 
and it propagates along a ring over a rotation period of the 
galaxy - thus the stellar age along the ring should show a gradient.   
No such gradient is expected in star formation triggered by a tidal event.
A detailed stellar age study is required to verify this result.

(6) The star forming regions appear to be correlated with the H{\sc i} distribution 
in five of the seven GLSB galaxies studied here. However little correlation 
between the NUV emission and H{\sc i} is visible in  UM~163.   
A burst of star formation might have exhausted high column density gas. 

(7) Since GLSB galaxies are not subject to intense tidal interactions 
\citep{bothun93}, the mean H{\sc i} surface densities have 
remained low -may be indicating the important role tidal interactions 
play in enhancing H{\sc i} surface densities and condensation into 
molecular gas in HSB galaxies.

(8) Finally, there appears to be no compelling reasons to support 
different formation scenarios for LSB and HSB spirals since the 
primordial quantities like galaxy masses (stellar disk + 
H{\sc i} content + dark matter content) and rotation velocities are 
statistically comparable.  The differences in star formation rates, 
mass of the central black hole and H{\sc i} surface densities are 
dynamic on shorter timescales and can be explained by the under-dense
environments that these LSB galaxies inhabit and the
comparatively dense environment that HSB galaxies inhabit.

\section*{Acknowledgments}

We thank the referee, Greg Bothun for insightful comments that 
improved the content of this paper. We thank the staff of the 
GMRT who made the observations possible.
The GMRT is operated by the National Centre for Radio Astrophysics
of the Tata Institute of Fundamental Research. NGK acknowledges the 
generous use of NASA's Astrophysics Data System, NASA/IPAC Extragalactic 
Database (NED),  Hyperleda database (http://leda.univ-lyon1.fr),  
Astrophysics arXivs, Wikipedia and Google search engine in this research. 
This publication makes use of data products from the 2MASS, which is a joint 
project of the University of Massachusetts and the Infrared Processing and
Analysis Center/California Institute of Technology, funded by the
National Aeronautics and Space Administration and the National
Science Foundation. The NASA/IPAC Extragalactic Data base (NED) both of 
which are operated by Jet Propulsion Laboratory, California Institute of 
Technology under contract with the National Aeronautics and
Space Administration.

\bibliographystyle{mn2e}
\bibliography{alka.et.al}

\end{document}